\documentclass[preprint,12pt]{elsarticle}

\usepackage{amssymb,float}
\journal{Annals of Physics}

\begin{document}

\begin{frontmatter}

%% Title, authors and addresses

%% use the tnoteref command within \title for footnotes;
%% use the tnotetext command for the associated footnote;
%% use the fnref command within \author or \address for footnotes;
%% use the fntext command for the associated footnote;
%% use the corref command within \author for corresponding author footnotes;
%% use the cortext command for the associated footnote;
%% use the ead command for the email address,
%% and the form \ead[url] for the home page:
%%
%% \title{Title\tnoteref{label1}}
%% \tnotetext[label1]{}
%% \ead[email]{luigi.coraggio@na.infn.it}
%% \ead[url]{home page}
%% \fntext[label2]{}
%% \address{Address\fnref{label3}}
%% \fntext[fn1]{}

\title{Effective shell-model hamiltonians from realistic
  nucleon-nucleon potentials within a perturbative approach}

%% use optional labels to link authors explicitly to addresses:
%% \author[label1,label2]{<author name>}
%% \address[label1]{<address>}
%% \address[label2]{<address>}

\author[label1]{L. Coraggio\corref{cor1}}
\cortext[cor1]{Corresponding author}
\ead{luigi.coraggio@na.infn.it}
\fntext[]{Phone +39 081 676272, Fax +39 081 676904}
\author[label1,label2]{A. Covello}
\author[label1]{A. Gargano}
\author[label1,label2]{N. Itaco}
\author[label3]{T. T. S. Kuo}
\address[label1]{Istituto Nazionale di Fisica Nucleare, \\
Complesso Universitario di Monte  S. Angelo, Via Cintia - I-80126 Napoli,
Italy}
\address[label2]{Dipartimento di Scienze Fisiche, Universit\`a
di Napoli Federico II, \\
Complesso Universitario di Monte  S. Angelo, Via Cintia - I-80126 Napoli,
Italy}
\address[label3]{Department of Physics, SUNY, Stony Brook, New York 11794}

\begin{abstract}
This paper discusses the derivation of an effective shell-model
hamiltonian starting from a realistic nucleon-nucleon potential by way
of perturbation theory.
More precisely, we present the state of the art of this approach when
the starting point is the perturbative expansion of the $\hat{Q}$-box
vertex function.
Questions arising from diagrammatics, intermediate-states and
order-by-order convergences, and their dependence on the chosen
nucleon-nucleon potential, are discussed in detail, and the
results of numerical applications for the $p$-shell model space
starting from chiral next-to-next-to-next-to-leading order
potentials are shown.
Moreover, an alternative graphical method to derive the effective
hamiltonian, based on the $\hat{Z}$-box vertex function recently
introduced by Suzuki {\it et al.}, is applied to the case of a
non-degenerate $(0+2)~\hbar \omega$ model space.
Finally, our shell-model results are compared with the exact ones
obtained from no-core shell-model calculations.
\end{abstract}

\begin{keyword}
%% keywords here, in the form: keyword \sep keyword

%% MSC codes here, in the form: \MSC code \sep code
%% or \MSC[2008] code \sep code (2000 is the default)
Nuclear shell model \\
Realistic effective interactions \\
$p$-shell nuclei
\end{keyword}

\end{frontmatter}

%%
%% Start line numbering here if you want
%%
% \linenumbers

%% main text
\section{Introduction}
\label{intro}
The shell model is the basic theoretical tool for the microscopic
description of nuclear structure.
As is well known, this model is based on the hypothesis that, as a
first approximation, each nucleon inside the nucleus moves independently
from the others, in a spherically symmetric mean field.
The nucleons then arrange themselves into groups of levels, the
``shells'', well separated from each other.

The shell model reduces the complex nuclear many-body problem to a
very simplified one, where only a few active nucleons (valence
nucleons) interact in a truncated model space spanned by a single
major shell above an inert core.

This scheme does not take into account neither the degrees of freedom
of the core nucleons nor the excitations of the valence nucleons into
the shells above the model space.
Actually, the physical observables must be calculated within an
effective theory, namely the shell-model hamiltonian has to take into
account all the degrees of freedom that are not considered explicitly.

The derivation of the effective shell-model hamiltonian may follow two
paths.
The first and very successful one is phenomenological, where the one-
and two-body components of the hamiltonian - the single-particle (SP)
energies and the residual interaction - contain free parameters that
are fixed to reproduce a set of experimental data.
This can be done either using an analytical expression for the
residual interaction with adjustable parameters, or treating the
hamiltonian matrix elements directly as free parameters (see
\cite{Elliott69a,Talmi03}).

The alternative approach to the derivation of the effective
shell-model hamiltonian is the microscopic one, where one starts from
a realistic nucleon-nucleon ($NN$) potential, and possibly a
three-nucleon one ($3N$), and derive the effective hamiltonian in the
framework of the many-body theory.
This means that when diagonalizing the shell-model hamiltonian in the
model space, its eigenvalues belong to the set of eigenvalues of the
full nuclear hamiltonian, defined in the whole Hilbert space.

The most successful way to derive a realistic effective shell-model
hamiltonian is rooted in the energy-independent linked-diagram
perturbation theory \cite{Kuo90}, which has been widely employed in
shell-model calculations during the last forty years (see review papers
\cite{Hjorth95,Coraggio09a}).
The core of this approach is the perturbative expansion of a vertex
function, the so-called $\hat{Q}$-box, as a collection of irreducible
valence-linked Goldstone diagrams.
The $\hat{Q}$-box is then employed to solve non-linear matrix
equations to derive the desired effective hamiltonian, which can be
done by way of iterative techniques \cite{Suzuki80}.
In all applications the Lee-Suzuki and Krenciglowa-Kuo iterative
methods have been commonly employed, both of them based on the
calculation of the derivatives of the $\hat{Q}$-box with respect to
the energy.

Recently, Suzuki and coworkers \cite{Suzuki11} have introduced an
alternative way to derive the shell-model effective hamiltonian, which
is a graphical method based on the introduction of another vertex
function, the so-called $\hat{Z}$-box, whose main advantages are that
only the first derivative of the $\hat{Q}$-box is needed, and that it
can be easily extended to the case of a non-degenerate model space.

In such a scenario, one should keep in mind that in modern nuclear
structure calculations it has been evidenced the role of $3N$ forces,
in particular for light nuclei with $A \leq 12$
\cite{Pieper01,Pieper05,Navratil07a}.
Within the framework of the shell-model effective hamiltonian theory,
the inclusion of $3N$ forces yields an effective hamiltonian that
consists of one- and two-body components, including also
core-polarization effects arising from the $3N$ force, and an effective
three-body interaction that should be explicitly considered in the
calculations.
The derivation of such a hamiltonian, even if not requiring any
refinement of the theory, is a very hard task, and up to the present a
shell-model effective hamiltonian derived treating on equal footing
both $NN$ and $3N$ forces has never been calculated.
In this connection it is worth to cite the work by Otsuka {\it et al.}
\cite{Otsuka10}, where only first-order contributions of the
normal-ordered two-body parts of $3N$ forces are taken explicitly into
account. 
On these grounds, we confine our study, without any loss of
generality, to effective shell-model hamiltonians derived starting
from purely two-body realistic potentials.

The aim of the present work is to try to assess the state of the art
of the derivation of the realistic shell-model effective hamiltonian
from the perturbative expansion of the $\hat{Q}$-box.
In particular, we shall discuss in detail the behavior of the
perturbative series, with respect to both the dimension of the space
of the intermediate states and the order-by-order convergence.

In this context, we will show the results of shell-model calculations
for $p$-shell nuclei starting from realistic $NN$ potentials based on
the chiral perturbation theory at next-to-next-to-next-to-leading
order (N$^3$LO) \cite{Entem03,Machleidt11,Coraggio07b}.
Moreover, we shall present results of calculations within the
$(0+2)~\hbar \omega$ $psd$ model space, using an effective hamiltonian
derived by the graphical $\hat{Z}$-box method.

The paper is organized as follows. 
In Section 2 we give an outline of the derivation of the shell-model
effective hamiltonian within a perturbative approach.
Section 3 is devoted to the discussion of the problematics concerning
the diagrammatics and convergence of the $\hat{Q}$-box perturbative
expansion.
This is done showing our results for the $p$-shell nuclei, using both
$0$ and $(0+2)~\hbar \omega$ model spaces.
In Section 4 we compare shell-model results obtained starting from the
N$^3$LO potential \cite{Entem03,Machleidt11} with the exact ones provided by the
{\em ab initio} no-core shell model \cite{Navratil04,Navratil07a}.
A summary and concluding remarks are given in Section 5.

\section{The shell-model effective hamiltonian}

In this section, we describe the formalism of the shell-model
effective hamiltonian theory.

Let us consider the Schr\"odinger equation for the $A$-nucleon system
in the whole Hilbert space:
\begin{equation}
H | \Psi_{\nu} \rangle  = E_{\nu} | \Psi_{\nu} \rangle \label{eq1}~~.
\end{equation}

As mentioned in the Introduction, in the frame of the shell-model
approach an auxiliary one-body potential $U$ is introduced to break
up the nuclear hamiltonian as the sum of an unperturbed one-body term
$H_0$, which describes the independent motion of the nucleons, and the
interaction hamiltonian $H_1$.

We then write $H=H_0+H_1$ with $H_0=\sum_{i=1}^A h_{0_i}= \sum_{i=1}^A
(T_i+U_i)$ and $H_1=\sum_{i<j}(V^{NN}_{ij}-U_i)$.

In the shell model, the nucleus is schematized as an inert core plus
$n$ interacting valence nucleons moving in the mean field $H_0$.
The SP potential $U$ generates an energy spectrum organized in
shells.
The large energy difference between shells enables to define the core
as the $A-n$ nucleons that fill completely the lowest shells.
The SP states accessible to the $n$ valence nucleons are then the
lowest in energy above the closed core.

Now, it is possible to define a reduced Hilbert space, the model space, 
in terms of a finite subset of $d$ eigenvectors of $H_0$
\begin{equation}
|\Phi_i \rangle = [ a^{\dagger}_1 a^{\dagger}_2 ~...~a^{\dagger}_n ]_i
| c \rangle ~~,
\end{equation}

\noindent
where $|c \rangle$ represents the inert core, the subscripts 1, 2,
..., $n$ denote the SP valence states and $i$ stands for all the other
quantum numbers that specify the state.

More explicitly, we define the projection operators $P$ and $Q=1-P$,
which project from the complete Hilbert space onto the model space and
its complementary space, respectively.
They satisfy the properties $P^2=P$, $Q^2=Q$, and $PQ=QP=0$.
In terms of eigenvectors of $H_0$, $P$ is defined as 
\begin{equation}
P= \sum_{i=1}^d | \Phi_i \rangle \langle \Phi_i |~~.
\end{equation}

Our aim is to reduce the eigenvalue problem of Eq. (\ref{eq1}) to the
model-space eigenvalue problem
\begin{equation}
H_{\rm eff} P | \Psi_{\alpha} \rangle  = E_{\alpha} P | \Psi_{\alpha}
\rangle \label{eq3}~~,
\end{equation}

\noindent
where $\alpha=1,..,d$ and $H_{\rm eff}$ is defined only in the model space.
In other words, we are looking for a new hamiltonian $\mathcal{H}$
that has the same eigenvalues of the $A$-nucleon system hamiltonian
$H$, and satisfies the decoupling equation between the model space $P$
and its complement $Q$:

\begin{equation}
Q \mathcal{H} P=0 ~~, \label{deceq1}
\end{equation}

\noindent
so that the desired effective hamiltonian is $H_{\rm eff}= P
\mathcal{H} P$.

The new hamiltonian $\mathcal{H}$ can be obtained by way of a
similarity transformation defined in the whole Hilbert space:

\begin{equation}
\mathcal{H}=X^{-1} H X ~~.
\end{equation}
\noindent

There is of course an infinite class of transformation operators
$X$ that satisfy the decoupling equation (\ref{deceq1}). 
Suzuki and Lee \cite{Suzuki80} considered an operator $X$ defined as
$X=e^{\omega}$.
Without loss of generality, they chose $\omega$ so as to satisfy the
following properties:

\begin{equation}
\omega= Q \omega P ~, 
\label{omegapro1}
\end{equation}
\begin{equation}
P \omega P= Q \omega Q = P \omega Q =0 ~. 
\label{omegapro2}
\end{equation}

\noindent
Eq. (\ref{omegapro1}) implies that 

\begin{equation}
\omega ^2 = \omega^3 = ~...~=0 ~. 
\label{omegapro3}
\end{equation}

The above equation enables to write $X=1+ \omega$, and consequently it
can be obtained

\begin{equation}
H_{\rm eff} = P \mathcal{H} P = PHP +PH Q \omega ~~.
\end{equation}

The operator $\omega$ may be obtained by solving the decoupling equation
(\ref{deceq1}), which may be rewritten in the following form

\begin{equation}
Q H P + Q H Q \omega - \omega P H P - \omega P H Q \omega = 0
~~. \label{deceq2} 
\end{equation}

The above matrix equation is non-linear and can be easily solved, once 
the hamiltonian $H$ is explicitly known in the whole Hilbert space.
However, this is not an easy task for nuclei with mass $A>2$, and has been
recently done only for light nuclei within the framework of {\em
  ab-initio} approaches \cite{Lisetskiy08}.

The standard approach to solve Eq. (\ref{deceq2}) in a shell-model
calculation is to introduce a vertex function, the $\hat{Q}$-box, that
can be then expressed as a perturbative expansion.

Let us consider our model space, which we assume to be degenerate:

\begin{equation}
P H_0 P =\epsilon_0 P~~. \label{dege}
\end{equation}

Then, taking into account the decoupling equation (\ref{deceq1}), the
effective hamiltonian $H^{\rm eff}_1=H_{\rm eff} - P H_0 P$ can be
written in terms of $\omega$

\begin{equation}
H^{\rm eff}_1 = P \mathcal{H} P - P H_0 P = P H_1 P + P H_1 Q \omega
~~. \label{eqqq}
\end{equation}

We now employ the above identity, the decoupling equation
(\ref{deceq2}), and the properties of $H_0$ and $H_1$, to obtain a
recursive equation for the effective hamiltonian $H^{\rm eff}_1$.

First, since $H_0$ is diagonal, we can write the following identity:

\begin{equation}
QHP= QH_1P + QH_0P = QH_1P~~.
\end{equation}

The decoupling equation (\ref{deceq2}) can then be rewritten in the
following form:

\begin{equation}
Q H_1 P + Q H Q \omega - \omega (P H_0 P + P H_1 P + P H_1 Q \omega) = 
Q H_1 P + QHQ \omega - \omega ( \epsilon_0 P + H_1^{\rm eff}) = 0
~~. \label{deceq3} 
\end{equation}

Using this form of the decoupling equation, we can write the following
identity for the operator $\omega$:

\begin{equation}
\omega = Q \frac{1}{\epsilon_0 -QHQ} Q H_1 P - Q \frac{1}{\epsilon_0
    -QHQ} \omega H^{\rm eff}_1~~. \label{omegaq}
\end{equation}

Finally, inserting Eq. (\ref{omegaq}) into the identity (\ref{eqqq})
that defines $H^{\rm eff}_1$, we obtain a recursive equation:

\begin{equation}
H^{\rm eff}_1 (\omega) = P H_1 P + P H_1 Q \frac{1}{\epsilon_0 - Q H Q} Q
  H_1 P - P H_1 Q \frac{1}{\epsilon_0 - Q H Q} \omega H^{\rm eff}_1
  (\omega) ~~. \label{eqsemifinal}
\end{equation}

We define the vertex function $\hat{Q}$-box by the following identity:

\begin{equation}
\hat{Q} (\epsilon) = P H_1 P + P H_1 Q \frac{1}{\epsilon - Q H Q} Q
H_1 P ~~, \label{qbox}
\end{equation}

\noindent
so that the recursive equation (\ref{eqsemifinal}) can be expressed as

\begin{equation}
H^{\rm eff}_1 (\omega) = \hat{Q}(\epsilon_0) - P H_1 Q \frac{1}{\epsilon_0
  - Q H Q} \omega H^{\rm eff}_1 (\omega) ~~. \label{eqfinal}
\end{equation}

Suzuki and Lee \cite{Suzuki80} suggested two possible iterative
techniques to solve Eq. (\ref{eqfinal}), both of them based on the
calculation of $\hat{Q}$-box derivatives.
In the next subsections, we shall briefly describe these two methods,
which have become known as the Krenciglowa-Kuo (KK) and the Lee-Suzuki
(LS) techniques.

\subsection{The Krenciglowa-Kuo iterative technique}
This iterative approach originates from the observation that
Eq. (\ref{eqfinal}), coupled with the recursive expression for the
operator $\omega$, Eq. (\ref{omegaq}), leads to the iterative
equation:

\begin{equation}
 H^{\rm eff}_1 (\omega_n) = 
\sum_{m=0}^{\infty} \left[-P H_1 Q \left( \frac{-1}{\epsilon_0 -QHQ}
  \right)^{m+1} QH_1P \right] \left[ H^{\rm eff}_1 (\omega_{n-1})
\right]^m ~~.
\label{eqa}
\end{equation}

It should be noted that the quantity inside the square brackets in
Eq. (\ref{eqa}), which from now on we write as $\hat{Q}_m(\epsilon_0)$,
is equal to:

\begin{equation}
\hat{Q}_m(\epsilon_0) = -P H_1 Q \left( \frac{-1}{\epsilon_0 -QHQ}
  \right)^{m+1} QH_1P = \frac{1}{m!} \left[ \frac{d^m \hat{Q} 
(\eta)} {d \eta^m} \right]_{\eta=\epsilon_0} ~~.
\label{eqb}
\end{equation}

This identity allows to rewrite Eq. (\ref{eqa}) in the final form:

\begin{equation}
 H^{\rm eff}_1 (\omega_n) =
\sum_{m=0}^{\infty} \frac{1}{m!} \left[ \frac{d^m \hat{Q} (\eta)} {d
    \eta^m} \right]_{\eta=\epsilon_0} \left[ H^{\rm eff}_1
  (\omega_{n-1}) \right]^m  = 
\sum_{m=0}^{\infty} \hat{Q}_m(\epsilon_0) \left[ H^{\rm eff}_1
  (\omega_{n-1}) \right]^m ~~.
\label{eqc}
\end{equation}

The above approach is known as the Krenciglowa-Kuo iterative method,
since, if we make the assumption $ H^{\rm eff}_1 (\omega_0)=\hat{Q}
(\epsilon_0)$, Eq. (\ref{eqc}) can be rewritten as

\begin{equation}
H^{\rm eff} = \sum_{i=0}^{\infty} F_i~~,
\end{equation}

\noindent
where

\begin{eqnarray}
F_0 & = & \hat{Q}(\epsilon_0)  \nonumber \\
F_1 & = & \hat{Q}_1(\epsilon_0)\hat{Q}(\epsilon_0)  \nonumber \\
F_2 & = & \hat{Q}_2(\epsilon_0)\hat{Q}(\epsilon_0)\hat{Q}(\epsilon_0) + 
\hat{Q}_1(\epsilon_0)\hat{Q}_1(\epsilon_0)\hat{Q}(\epsilon_0)  \nonumber \\
~~ & ... & ~~ 
\end{eqnarray}

This is the well-known folded-diagram expansion of the effective
hamiltonian as introduced by Kuo and Krenciglowa in
\cite{Krenciglowa74}, where the following operatorial identity has
been demonstrated:

\begin{equation}
\hat{Q}_1\hat{Q}= - \hat{Q} \int \hat{Q}~~,
\end{equation}

\noindent
the integral sign representing the so-called folding operation
\cite{Brandow67}.

\subsection{The Lee-Suzuki iterative technique}
Suzuki and Lee \cite{Suzuki80} suggested another iterative technique,
which can be introduced by rearranging Eq. (\ref{eqfinal}) so as to have
an explicit expression of the effective hamiltonian $H^{\rm eff}_1$ in
terms of the operators $\omega$ and $\hat{Q}$:

\begin{equation} 
H^{\rm eff}_1 (\omega) = \left( 1 + P H_1 Q \frac{1}{\epsilon_0 - Q H Q}
  \omega \right)^{-1} \hat{Q} (\epsilon_0) ~~.
\end{equation}

The above equation can be rewritten in an iterative form:

\begin{equation} 
H^{\rm eff}_1 (\omega_n) = \left( 1 + P H_1 Q
\frac{1}{\epsilon_0 - Q H Q} \omega_{n-1} \right)^{-1} \hat{Q}
(\epsilon_0)~~,
\end{equation}

\noindent
and similary for the recursive equation (\ref{omegaq}):

\begin{equation} 
\omega_n = Q \frac{1}{\epsilon_0 -QHQ} Q H_1 P - Q \frac{1}{\epsilon_0
    -QHQ} \omega_{n-1}H^{\rm eff}_1(\omega_n) ~~.
\end{equation}

The iterative procedure can start from the choice $\omega_0=0$, so we
can write:

\begin{eqnarray}
H^{\rm eff}_1 (\omega_1) & = & \hat{Q} (\epsilon_0) \nonumber \\
\omega_1 & = &  Q \frac{1} {\epsilon_0 -Q H Q} Q H_1 P ~~. \nonumber
\end{eqnarray}

By doing some algebra, it is possible to demonstrate:

\begin{eqnarray}
 \hat{Q}_1 (\epsilon_0) & = & - P H_1 Q \frac{1}{\epsilon_0 - QHQ} Q
\frac{1}{\epsilon_0 - QHQ} Q H_1 P \nonumber \\
~ & = & - P H_1 Q \frac{1}{\epsilon_0 - QHQ}
\omega_1 ~~,
\end{eqnarray}

\noindent
so that, for the next iteration $n=2$, one has:

\begin{eqnarray}
H^{\rm eff}_1 (\omega_2) & = & \left( 1 + P H_1 \frac{1} {\epsilon_0 -
    QHQ} \omega_1 \right)^{-1} \hat{Q}(\epsilon_0) =  \nonumber \\
~ & = & \frac{1}{1 - \hat{Q}_1(\epsilon_0)} \hat{Q} (\epsilon_0) \nonumber \\
\omega_2 & = & Q \frac{1}{\epsilon_0 - QHQ} Q H_1 P -  Q
\frac{1}{\epsilon_0 -QHQ} \omega_{1}H^{\rm eff}_1(\omega_2)  ~~.
\end{eqnarray}

Finally, the iterative form of the equation for the effective
hamiltonian within the LS approach reads:

\begin{equation}
H^{\rm eff}_1 (\omega_n) = \left[
1 - \hat{Q}_1 (\epsilon_0) \sum_{m=2}^{n-1} \hat{Q}_m (\epsilon_0)
\prod_{k=n-m+1}^{n-1} H^{\rm eff}_1 (\omega_k)
\right]^{-1} \hat{Q} (\epsilon_0)  ~~. \label{eqLS}
\end{equation}

The KK and LS techniques to solve the decoupling equation do not
necessarily provide the same effective hamiltonian.
In Ref. \cite{Suzuki80} it has been shown that, when converging, the
KK iterative procedure provides an effective hamiltonian whose
eigenstates have the largest model space overlap, while the effective
hamiltonian obtained with the LS one has eigenvalues that are the
lowest in energy.

It is worth noting that both procedures are constrained to choose an
unperturbed hamiltonian $H_0$ whose eigenstates belonging to the model
space are degenerate in energy.
In Ref. \cite{Kuo95} an alternative approach to the standard KK and LS
procedures has been proposed, which extends these methods to the
non-degenerate case by introducing multi-energy $\hat{Q}$-boxes.
However, this approach has proved to be quite complicated for practical
applications, the only one appeared in the literature being that of
Ref. \cite{Coraggio05c}.

In the following subsection, another method to calculate the
shell-model effective hamiltonian, recently derived by Suzuki {\em et
  al.} \cite{Suzuki11}, will be outlined.

\subsection{The $\hat{Z} (\epsilon)$ vertex function}
\label{zbox}
From inspection of Eq. (\ref{qbox}), which defines the
$\hat{Q}$-box, it can be seen that when $\epsilon$ approaches one of
the eigenvalues of $QHQ$, $\hat{Q}(\epsilon)$ has some poles, which
can induce instabilities in the numerical derivation.

To overcome such these difficulties, Suzuki {\em et al.}
\cite{Suzuki11} have recently introduced an alternative vertex
function $\hat{Z}(\epsilon)$ that is defined in terms of
$\hat{Q}(\epsilon)$ and its first derivative:

\begin{equation}
\label{eq:z-box}
 \hat Z(\epsilon) \equiv \frac{1}{{1 - \hat Q_1 (\epsilon)}}\left[
   {\hat Q(\epsilon) - \hat Q_1 (\epsilon) (\epsilon - \epsilon_0 ) P
   } \right]~~.
 \end{equation}

In Ref. \cite{Suzuki11} it has been shown that $\hat Z(\epsilon)$
satisfies the following equation

\begin{equation}
\left[ \epsilon_0  + \hat Z(E_\alpha) \right] P | \Psi_{\alpha}
\rangle  = E_{\alpha} P | \Psi_{\alpha} \rangle ~~~~~~~~~~(\alpha=1,..,d)~~, 
\end{equation}

\noindent
which means that $H^{\rm eff}_1$ may be obtained calculating the
$\hat{Z}$-box for those values of the energy, determined
self-consistently, that correspond to the ``true'' eigenvalues
${E_\alpha}$. 

To obtain the ${E_\alpha}$, let us now consider the eigenvalue problem 
\begin{equation} 
 \left[ {\epsilon_0  + \hat Z(\epsilon)} \right] | {\phi_k
 }\rangle=F_{k}(\epsilon) |{\phi_k}\rangle~~,~~~~~~~~~~ (k = 1,2,\cdots,d)~~, 
\label{eq-Z-eigenvalue}
 \end{equation}

\noindent
where $F_{k}(\epsilon)$ are $d$ eigenvalues that depend on $\epsilon$.
The true eigenvalues ${E_\alpha}$ may be determined by solving the
following $d$ equations

\begin{equation}
\label{eq:EFK}
\epsilon=F_k(\epsilon), \,\,\,\,(k=1,2,\cdots ,d)~~.
\end{equation}
 
Before discussing how to solve equations
(\ref{eq-Z-eigenvalue},\ref{eq:EFK}), let us point out some
interesting properties of $\hat Z(\epsilon)$ and of the associated
functions $F_k(\epsilon)$.

In the neighborhood of the poles of $\hat Q(\epsilon)$, the behavior
of $\hat Z(\epsilon)$ is dominated by $\hat Q_1(\epsilon)$ and so it
can be written $\hat{Z}(\epsilon) \approx (\epsilon - \epsilon_0)P$.
This means that $\hat{Z}(\epsilon)$ has no poles and therefore the
functions $F_k(\epsilon)$ are continuous and differentiable for any
value of $\epsilon$. 

Eqs. (\ref{eq:EFK}) may have spurious solutions, i.e. solutions that
do not correspond to the true eigenvalues ${E_\alpha}$.
However, in Ref. \cite{Suzuki11} it has been shown that the energy
derivative of $F_k(\epsilon)$ becomes zero at $\epsilon = E_\alpha$,
so the study of this derivative gives a criterion to get rid of the
spurious solutions.
In order to solve Eqs. (\ref{eq-Z-eigenvalue}) and (\ref{eq:EFK}), so
as to derive the effective interaction, one may resort to both
iterative and non-iterative methods.

In the present paper we have employed a graphical non-iterative method
to solve Eqs. (\ref{eq:EFK}), which we shall now describe.

As shown previously, the $F_k(\epsilon)$'s are continuous functions of
the energy, therefore one of the well-known algorithms to solve
nonlinear equations may be employed to determine the solutions of
Eqs. (\ref{eq:EFK}) as the intersections of the graphs $y = \epsilon$
and $y = F_k(\epsilon)$.

More precisely, if we define the functions $f_k(\epsilon)$ as
$f_k(\epsilon) = F_k(\epsilon) - \epsilon$, the solutions of
Eqs. (\ref{eq:EFK}) can obtained by finding the roots of the equations
$f_k(\epsilon) = 0$.
To this end, from inspection of the graphs $y = \epsilon$ and $y =
F_k(\epsilon)$, we determine for each intersection a small surrounding
interval $[\epsilon_a , \epsilon_b] $ such that $f_k(\epsilon_a)
f_k(\epsilon_b) < 0 $. 
The assumption that $f_k(\epsilon)$ is monotone in this interval
implies the existence of a unique root, that we can determine very
accurately employing the secant method algorithm (see for instance
Ref. \cite{recipes_for}).

Once we have determined the true eigenvalues ${E_\alpha}$, the
effective hamiltonian $H^{\rm eff}_1$ is built up as
\begin{equation}
\label{eq:REKK}
 H^{\rm eff}_1 = 
 \sum\limits_{\alpha = 1}^d {\hat Z} (E_\alpha )|{\phi_\alpha}\rangle 
 \langle {\tilde\phi_\alpha}| ~~,
\end{equation}

\noindent
where $|{\phi_\alpha}\rangle $ is the eigenvector obtained from
Eq. (\ref{eq-Z-eigenvalue}) while  $\langle {\tilde\phi_\alpha}|$ is
the correspondent biorthogonal state ($\langle{\tilde\phi_{\alpha} |\phi_{\alpha'}
}\rangle  = \delta_{\alpha \alpha'}$).

In conclusion, it is worth pointing out that in the previous
discussion we have considered the case of a degenerate unperturbed
model space (i.e., $ P H_0 P =\epsilon_0 P $).
However, the above formalism can be easily generalized to the
non-degenerate case replacing $\epsilon_0 P $ with $ P H_0 P $ in
Eqs. (\ref{eq:z-box}) and (\ref{eq-Z-eigenvalue}).

\section{The perturbative approach to the shell-model $H^{\rm eff}$}
The present section is devoted to discuss details and problems of the
derivation of a shell-model effective hamiltonian by way of the
perturbative approach.
More precisely, the $\hat{Q}$-box is calculated perturbatively and
then employed to derive the effective hamiltonian within the KK,
LS, or $\hat{Z}$-box graphical approaches.
It should be pointed out that the above techniques lead to effective
hamiltonians whose matrix elements differ at most by a few keV when
considering the KK and LS methods, while when using the $\hat{Z}$-box
method the differences do not exceed 80 keV.
This can be seen from inspection of Tables \ref{tableTBMEN3LO} and
\ref{tableTBMEN3LOW} that can be found in \ref{effint}.

\subsection{The diagrammatic expansion of the $\hat{Q}$-box}
\label{qboxsec}
The methods presented in the previous subsections are based on the
calculation of the $\hat{Q}$-box function:

\begin{equation}
\hat{Q} (\epsilon) = P H_1 P + P H_1 Q \frac{1}{\epsilon - Q H Q} Q
H_1 P ~~. \nonumber
\end{equation}

The term $1/(\epsilon - Q H Q)$ can be expanded as a power series

\begin{equation}
\frac{1}{\epsilon - Q H Q} = \sum_{n=0}^{\infty} \frac{1}{\epsilon -Q
  H_0 Q} \left( \frac{Q H_1 Q}{\epsilon -Q H_0 Q} \right)^{n} ~~,
\end{equation}

\noindent
giving rise to a perturbative expression for the $\hat{Q}$-box.
The diagrammatic representation of this perturbative expansion is a
collection of diagrams that have at least one $H_1$-vertex, are
irreducible (i.e., with at least one line not belonging to the model
space between two successive vertices), and valence linked (i.e., are
linked to at least one external valence line) \cite{Kuo71}.
 
Currently, realistic shell-model effective hamiltonians are derived
for systems with one and two valence nucleons.
The former provides the theoretical effective SP energies, while the
two-body residual interaction $V^{\rm eff}$ is obtained from the
$H^{\rm eff}$ of the two-valence-nucleon system using a subtraction
procedure \cite{Shurpin83}.
Up-to-date applications include in the $\hat{Q}$-box at most Goldstone
diagrams up to third order in $H_1$, which take into account up to
$3p-2h$ excitations for the one valence-nucleon system, and up to
$4p-2h$ excitations for the two valence-nucleon system.
A comprehensive work concerning the evaluation of the linked Goldstone
diagrams in an angular momentum coupled representation may be found in
Ref. \cite{Kuo81}.

\begin{figure}[H]
\begin{center}
\includegraphics[scale=0.4,angle=0]{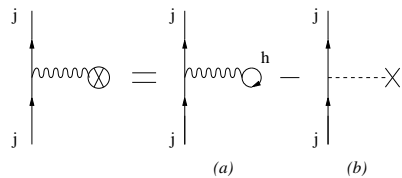}
\caption{One-body first-order diagram. Graph ($a$) is the so-called
  self-energy diagram. Graph ($b$) represents the matrix element of
  the harmonic oscillator potential $U=\frac{1}{2}m \omega^2 r^2$.}
\label{Sbox1}
\end{center}
\end{figure}

The one-body $\hat{Q}$-box diagram at first order in $H_1$ is reported in
Fig. \ref{Sbox1}.
All other $\hat{Q}$-box diagrams up to third order in $H_1$ can be
found in \ref{graphs}, and are reported in Fig. \ref{Sbox2},
\ref{Sbox3} (one-body diagrams), and in Fig. \ref{Qbox2},
\ref{Qbox3a}, \ref{Qbox3b}, \ref{Qbox3c} (two-body diagrams).

From inspection of Fig. \ref{Sbox1}, it can be seen that the
first-order one-body diagram is composed of the so-called self-energy
diagram ($V$-insertion diagram) minus the auxiliary potential
$U$-insertion.
The $U$-insertion diagrams arise in the perturbative expansion owing
to the presence of the $-U$ term in $H_1$.

The $(V-U)$-insertion diagrams turn out to be identically zero when
employing a self-consistent Hartree-Fock (HF) auxiliary potential
\cite{Coraggio05c}.
It is worth noting that in most applications the standard choice for the
auxiliary potential is the harmonic-oscillator (HO) one, and that the
$(V-U)$-insertion diagrams are neglected, assuming that the differences
between the HO and the HF single-particle wavefunctions are negligible.
In Subsection \ref{hfinsertions}, we shall discuss about the
contribution of these terms, comparing different effective
hamiltonians derived starting from $\hat{Q}$-boxes with and without
$(V-U)$-insertion diagrams.

As we pointed out before, in the existing literature the effective
hamiltonians are derived taking into account in the $\hat{Q}$-box at
most diagrams up to the third order, being computationally prohibitive
to go to higher-order.

In order to have a better estimate of the value to which the
perturbation series should converge, it is helpful to resort to
the Pad\`e approximant theory \cite{Baker70,Ayoub79}, and to calculate
the Pad\`e approximant $[2|1]$ of the $\hat{Q}$-box, as suggested in
\cite{Hoffmann76}:

\begin{equation}
[2|1] = V_{Qbox}^0 + V_{Qbox}^1 +V_{Qbox}^2
(1-(V_{Qbox}^2)^{-1}V_{Qbox}^{3})^{-1}~~,
\label{padeq}
\end{equation}

\noindent
where $V_{Qbox}^n$ is the square non-singular matrix representing the
$n$th-order contribution to the $\hat{Q}$-box in the perturbative
expansion.

In the following subsections we show results which we have obtained
calculating $H^{\rm eff}$ for two valence nucleons outside $^{4}$He
doubly-closed core in the $p$-shell model space, and using two
potentials based on the chiral perturbation theory at
next-to-next-to-next-to-leading order.
One is the well-known N$^3$LO potential derived by Entem and
Machleidt \cite{Entem03,Machleidt11}, that is characterized by a smooth gaussian
cutoff around 2.5 ${\rm fm}^{-1}$.
The other potential, dubbed N$^3$LOW \cite{Coraggio07b}, has a
sharp cutoff with a smaller value $\Lambda=2.1~{\rm fm}^{-1}$.
In both cases, for protons the Coulomb force has been explicitly added.

\subsection{Convergence with respect to the intermediate-state space}
The $Q$ space that enters the definition of the $\hat{Q}$-box in
Eq. (\ref{qbox}) is infinite, representing the complement of the model
space in the whole Hilbert space.
This implies that in the calculation of the diagrams composing the
$\hat{Q}$-box one should perform an infinite sum over the intermediate
states between successive vertices.
This is unfeasible, so the space of the intermediate states has
obviously to be truncated.
A well established procedure is to introduce an energy truncation,
i.e. the intermediate states whose unperturbed excitation energy is
greater than a fixed value $E_{max}$ are disregarded. 
$E_{max}$ has to be chosen sufficiently large to ensure that the
results are almost independent from its value. 

In this regard, it is appropriate to mention the papers by Vary {\em
  et al.} \cite{Vary73}, Kung {\em et al.} \cite{Kung79}, and
Sommermann {\em et al.} \cite{Sommermann81}, where the convergence
rate of the sum over intermediate-particle states in the second-order
core polarization contribution to the effective shell-model
interaction was studied, using realistic potentials renormalized by
way of the Brueckner theory \cite{Brueckner54}.

\begin{figure}[H]
\begin{center}
\includegraphics[scale=0.35,angle=0]{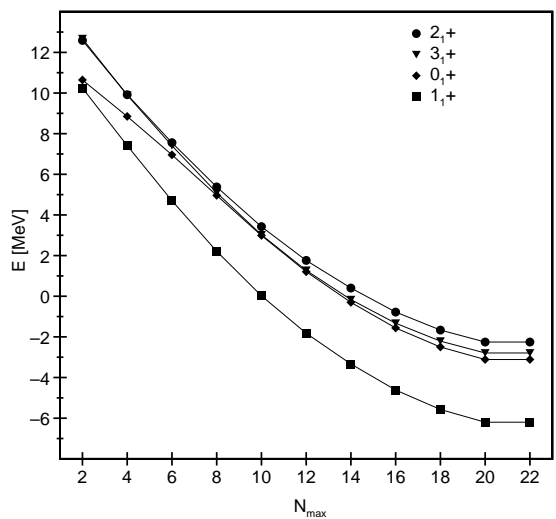}
\caption{Theoretical energies of $^{6}$Li yrast states relative to
  $^4$He, obtained with the N$^3$LO potential, as a function of
  $N_{\rm max}$ (see text for details).}
\label{intN3LO}
\end{center}
\end{figure}

\begin{figure}[H]
\begin{center}
\includegraphics[scale=0.35,angle=0]{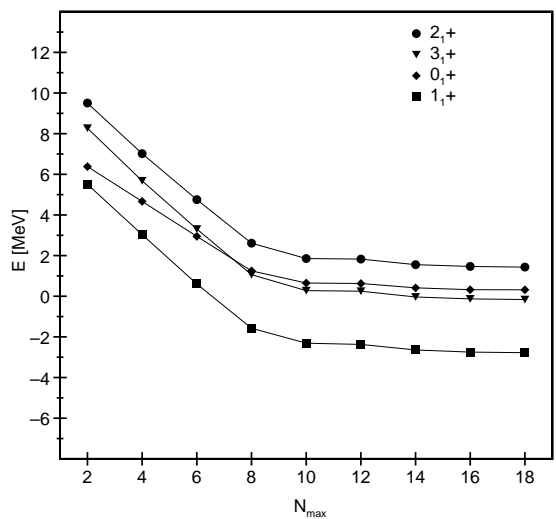}
\caption{Same as in Fig. \ref{intN3LO} for N$^3$LOW.}
\label{intN3LOW}
\end{center}
\end{figure}

In Figs. \ref{intN3LO} and \ref{intN3LOW}, the theoretical energies
of the yrast states in $^6$Li relative to $^4$He are reported as a
function of $E_{max}$, expressed in terms of the number of oscillator
quanta $N_{max}$.
The energies reported in Figs. \ref{intN3LO},\ref{intN3LOW} have been
calculated using an effective hamiltonian derived from the N$^3$LO and
N$^3$LOW potentials respectively, including in the $\hat{Q}$-box
diagrams up to the second order in $H_1$. 

From inspection of Fig. \ref{intN3LO}, it can be seen that
convergence is reached with the N$^3$LO potential when including
intermediate states whose unperturbed excitation energy is less than
$E_{max}=20~\hbar \omega$, with $\hbar \omega=19$ MeV.
This value of the harmonic oscillator parameter is close to the one
provided by the expression \cite{Blomqvist68} $\hbar \omega =
45A^{-1/3}-25A^{-2/3}$ for $A=4$.

A faster convergence can be obtained using the N$^{3}$LOW potential,
since, as shown in Fig. \ref{intN3LOW}, the energies are practically
stable from $E_{max}=10~\hbar \omega$ on.
This is related to the fact that the two potentials are characterized
by largely different cutoffs, N$^3$LOW being a low-momentum potential.

In most applications a subtraction procedure \cite{Shurpin83} is used
so as to retain from $H^{\rm eff}_1$ only the effective two-body
interaction $V^{\rm eff}$, while the SP energies are taken from
experiment.

We have therefore found it worthwhile to study the convergence
properties of $V^{\rm eff}$ when enlarging the space of intermediate
states.
To this end, we have kept fixed the set of SP energies, using the
one obtained from the effective hamiltonians with the largest number
of intermediate states, and calculated the energies of the yrast
states in $^6$Li relative to $^4$He, using  effective two-body
interactions $V^{\rm eff}$ that correspond to different values of
$N_{\rm max}$.

\begin{figure}[H]
\begin{center}
\includegraphics[scale=0.35,angle=0]{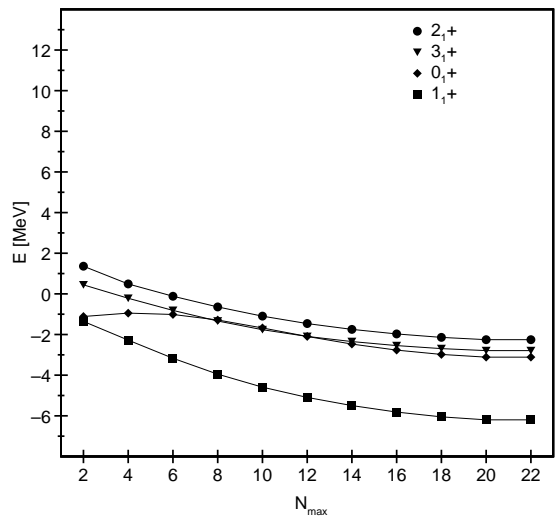}
\caption{Theoretical energies of $^{6}$Li yrast states relative to
  $^4$He, obtained with the N$^3$LO potential, as a function of
  $N_{\rm max}$ (see text for details).}
\label{intVN3LO}
\end{center}
\end{figure}

\begin{figure}[h]
\begin{center}
\includegraphics[scale=0.35,angle=0]{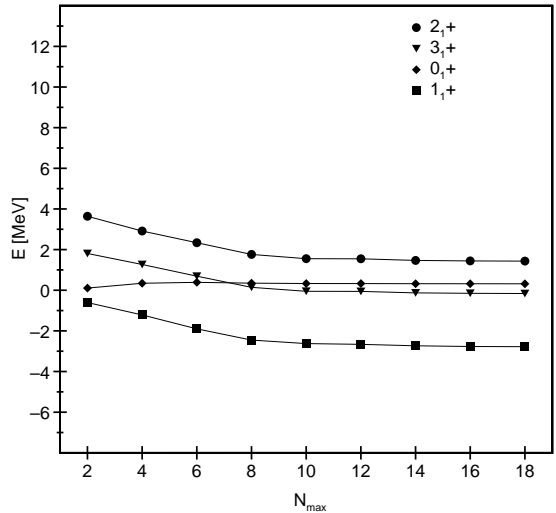}
\caption{Same as in Fig. \ref{intVN3LO} for N$^3$LOW.}
\label{intVN3LOW}
\end{center}
\end{figure}

The results are shown in Figs. \ref{intVN3LO} and \ref{intVN3LOW} for
the N$^3$LO and N$^3$LOW potentials respectively.

Comparing these figures with the previous Figs. \ref{intN3LO} and
\ref{intN3LOW} it can be seen that a faster convergence is obtained.
This may support the choice of smaller intermediate state spaces when
the SP energies are not derived theoretically but taken, for instance,
from experiment.

From now on, our calculations will refer to an intermediate state
space with $N_{max}=22$ and $N_{max}=18$ for the N$^3$LO and N$^3$LOW
potentials, respectively.

It is worth noting that the use of a second-order $\hat{Q}$-box does
not imply a loss of generality of the foregoing discussion, since both
at second and third order in the $\hat{Q}$-box expansion the $Q$ space
is spanned by the same particle-particle, 3-particle 1-hole, and
4-particle 2-hole excitations.
Therefore the conclusions drawn previously still hold for an effective
hamiltonian derived including diagrams up to third order in the
$\hat{Q}$-box expansion.

Concluding this subsection, we would like to point out that an
alternative approach to the study of the convergence with respect to
the intermediate-state space could be to sum at all orders the ladder
diagrams of the perturbative expansion by calculating the Brueckner
reaction-matrix $G$ of the N$^3$LO potential.

However, this is not an easy task, since to manage the operator
$Q_{2p}=1-P$ in the integral equation that defines the $G$ matrix,

\begin{equation}
G(\omega) = V_{NN} - V_{NN} Q_{2p} \frac{1}{\omega - H_0} Q_{2p}
G(\omega)~~, \label{gmat1}
\end{equation}

\noindent
would mean to handle matrix elements of the $NN$ potential $V_{NN}$ in
a very large space spanned by the HO wavefunctions.

A way to simplify the calculation of the $G$ matrix is to resort to
the so-called $G_T$ matrix \cite{Krenciglowa76} for which plane waves
are used as intermediate states:

\begin{equation}
G_T(\omega) = V_{NN} - V_{NN} Q_{2p} \frac{1}{\omega - Q_{2p} T Q_{2p}} Q_{2p}
G_T(\omega)~~. \label{gtmat1}
\end{equation}

A main advantage of the $G_T$ matrix is that an exact treatment of the
projection operator $Q_{2p}$ can be done employing the Tsai-Kuo method
\cite{Tsai72}.

\begin{figure}[H]
\begin{center}
\includegraphics[scale=0.30,angle=0]{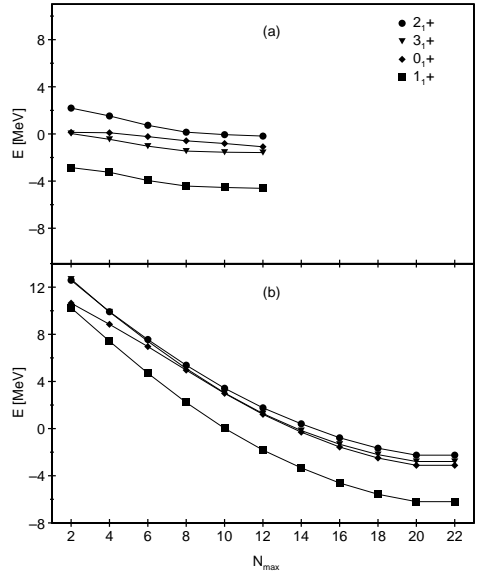}
\caption{Theoretical energies of $^{6}$Li yrast states relative to
  $^4$He, obtained with the $G_T$ matrix derived from the N$^3$LO
  potential, as a function of $N_{\rm max}$ (see text for details).}
\label{GintN3LO}
\end{center}
\end{figure}

By renormalizing the N$^3$LO potential with the $G_T$ matrix
procedure, we have calculated the energies of the yrast states in
$^6$Li relative to $^4$He, using effective hamiltonians derived at the
second order in $H_1$, as a function of $N_{max}$.
The energies are reported in Fig. \ref{GintN3LO} (case (a)), and
compared with the results previously shown in Fig. \ref{intN3LO} (case
(b)).

From inspection of Fig. \ref{GintN3LO}, it is evident that the
renormalization of the N$^3$LO potential strongly reduce the
dependence on the number of intermediate states of the perturbative
expansion of $H_{\rm eff}$, and consequently a faster convergence with
respect to $N_{max}$ is reached.
One has to keep in mind, however, that, since the Brueckner reaction
matrix is energy dependent, the results (a) are slightly dependent on
the choice of the starting energy $\epsilon_0$ in Eq. (\ref{eqfinal}).

\subsection{Order-by-order convergence}
Now, it is time to focus our attention on the dependence of the
effective hamiltonian on the order at which the perturbative expansion
of the $\hat{Q}$-box is arrested.

Historically, this problem was faced first in Ref. \cite{Barrett70},
where some selected third-order and few fourth-order terms in the $G$
matrix were calculated, and the convergence order-by-order of the
perturbation series was investigated.

As mentioned in Subsec. \ref{qboxsec}, we have derived effective
hamiltonians using $\hat{Q}$-boxes at second order ($H^{\rm eff}_{\rm
  2nd}$) and third order ($H^{\rm eff}_{\rm 3rd}$) in perturbation
theory.
Besides, we have also derived effective hamiltonians calculating the
Pad\`e approximant $[2|1]$ \cite{Baker70,Ayoub79} of the $\hat{Q}$-box
($H^{\rm eff}_{\mbox{\tiny{Pad\`e}}}$).
The matrix elements of the latter can be found in the Tables in
\ref{effint}.

\begin{figure}[H]
\begin{center}
\includegraphics[scale=0.34,angle=90]{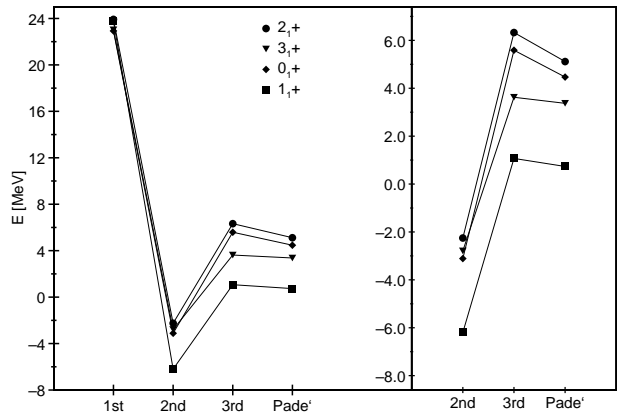}
\caption{Theoretical energies of $^{6}$Li yrast states relative to
  $^4$He, obtained with $H^{\rm eff}_{\rm 1st}$, $H^{\rm eff}_{\rm
    2nd}$, $H^{\rm eff}_{\rm 3rd}$, and $H^{\rm
    eff}_{\mbox{\tiny{Pad\`e}}}$ derived from the N$^3$LO potential
  (see text for details). In the right side of the figure, where an
  expanded scale is adopted, the $H^{\rm eff}_{\rm 1st}$ results are
  omitted.}
\label{N3LO3rd}
\end{center}
\end{figure}

\begin{figure}[H]
\begin{center}
\includegraphics[scale=0.34,angle=90]{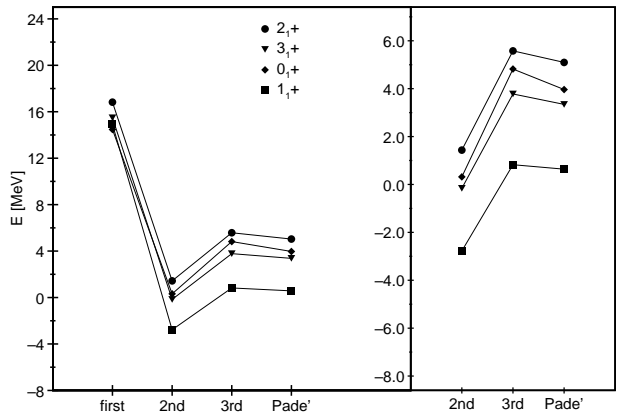}
\caption{Same as in Fig. \ref{N3LO3rd} for N$^3$LOW.}
\label{N3LOW3rd}
\end{center}
\end{figure}

Actually, in the definiton of the Pad\`e approximant $[2|1]$ (see
Eq. (\ref{padeq})) also the contribution at first order in $H_1$ comes
into play, so it is worth to consider also the results obtained
using only the diagram of Fig. \ref{Sbox1} for the one-body
hamiltonian, plus the $NN$ bare potential ($H^{\rm eff}_{\rm 1st}$).

In Figs. \ref{N3LO3rd} and \ref{N3LOW3rd}, we report the energies of
$^{6}$Li yrast states with respect to $^{4}$He, obtained with the N$^3$LO
and N$^3$LOW potentials, respectively, and calculated with $H^{\rm
  eff}_{\rm 1st}$, $H^{\rm  eff}_{\rm 2nd}$, $H^{\rm eff}_{\rm 3rd}$,
and $H^{\rm eff}_{\mbox{ \tiny{Pad\`e}}}$.

The large difference between the results with $H^{\rm eff}_{\rm 1st}$
and those obtained taking into account the correlations with the core
and the shells above the model space ($H^{\rm eff}_{\rm 2nd}$,
$H^{\rm eff}_{\rm 3rd}$) evidences the poor description provided by
the bare $NN$ potential without any renormalization due to long-range
correlations.

Comparing the results shown in the right sides of both figures, we
observe that the differences in the energies obtained with $H^{\rm
  eff}_{\rm 2nd}$ and $H^{\rm eff}_{\rm 3rd}$ are less notable using
the low-momentum N$^3$LOW potential.
This traces back to the fact that the N$^3$LO potential, even though
its gaussian cutoff is around $2.5~{\rm fm}^{-1}$, exhibits some
repulsion in the intermediate range, while the N$^3$LOW potential is a
real low-momentum potential.

\begin{figure}[H]
\begin{center}
\includegraphics[scale=0.35,angle=0]{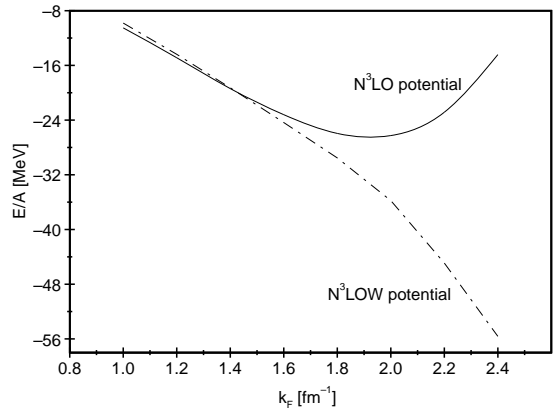}
\caption{Energy per nucleon in symmetric nuclear matter obtained from
  a Brueckner-Hartree-Fock calculation with N$^3$LO (smooth curve) and
  N$^3$LOW (dot-dashed curve).}
\label{nuclearmatter}
\end{center}
\end{figure}

This is testified by the fact that in nuclear matter the N$^3$LO
potentials saturates, while the N$^3$LOW potential does not; in
Fig. \ref{nuclearmatter} we report the energy per nucleon in symmetric
nuclear matter obtained from a Brueckner-Hartree-Fock calculation.
Obviously, the larger repulsive components of the N$^3$LO potential
influence negatively the order-by-order convergence.

It is worth pointing out, however, that both potentials lead to results
obtained with $H^{\rm eff}_{\mbox{\tiny{Pad\`e}}}$ that are very close to those
with $H^{\rm eff}_{\rm 3rd}$, thus supporting the hypothesis of a weak
dependence of the results on higher-order $\hat{Q}$-box perturbative
terms.

On the above grounds, in the following discussions we will employ only
effective hamiltonians derived by calculating the Pad\`e approximant
$[2|1]$ of the $\hat{Q}$-box.

Finally, it is appropriate to draw attention to the papers by
Hjorth-Jensen {\em et al.}
\cite{Hjorth92a,Hjorth92b,Hjorth92c,Hjorth96} where, within the
framework of the folded-diagram theory, the inclusion of the
third-order diagrams in the $G$-matrix in the calculation of the $\hat
Q$-box was studied and the order-by-order convergence of the effective
interaction was examined.
The main finding of Ref. \cite{Hjorth96} was that the effects of
third-order contributions in the $T=1$ channel are almost negligible.

\subsection{Dependence on the harmonic-oscillator parameter}
\label{hfinsertions}
As we have mentioned before, the shell-model effective hamiltonian is
derived using the harmonic-oscillator potential as the auxiliary
potential $U$.

Since we truncate the number of intermediate states and arrest the
perturbative expansion of the $\hat{Q}$-box at a certain order, this
introduces a dependence on the value of the harmonic-oscillator
parameter.

\begin{figure}[H]
\begin{center}
\includegraphics[scale=0.35,angle=90]{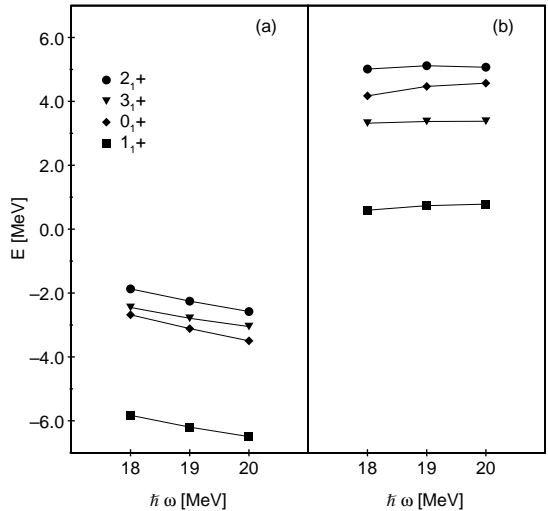}
\caption{Theoretical energies of $^{6}$Li yrast states relative to
  $^4$He, obtained with the N$^3$LO potential, as a function of $\hbar
  \omega$ (see text for details).}
\label{hf1}
\end{center}
\end{figure}

In order to study this dependence, we have derived from the N$^{3}$LO
potential three effective hamiltonians using $\hbar \omega$= 18, 19,
and 20 MeV, respectively.

In Fig. \ref{hf1}, the theoretical energies of the yrast states in
$^6$Li are reported as a function of $\hbar \omega$; case (a) refers
to effective hamiltonians derived including in the $\hat{Q}$-boxes
diagrams up to second order in perturbation theory.
Case (b) refers to effective hamiltonians derived including all
third-order diagrams in the $\hat{Q}$-box, and then calculating its
Pad\`e approximant $[2|1]$.

From inspection of Fig. \ref{hf1}, it can be observed that while
the eigenvalues of the second-order effective hamiltonians retain a
significative dependence on the harmonic-oscillator parameter, the
effective hamiltonians derived calculating the Pad\`e approximant
$[2|1]$ of the $\hat{Q}$-box are far less dependent on $\hbar \omega$.
It is worth to note, however, that in case (a) the relative spectra of
$^{6}$Li are almost independent of $\hbar \omega$.

\begin{figure}[H]
\begin{center}
\includegraphics[scale=0.35,angle=90]{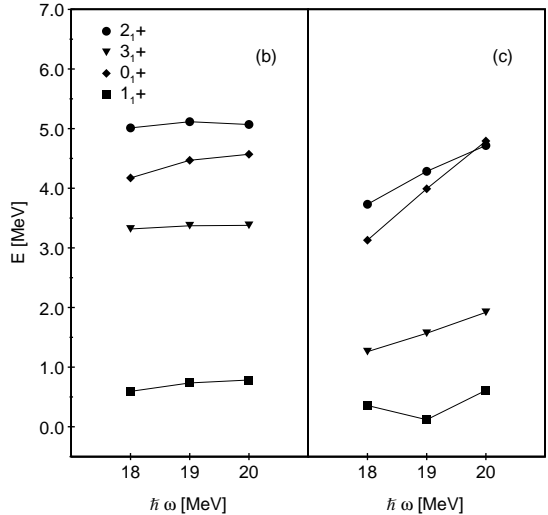}
\caption{Same as in Fig. \ref{hf1} (see text for details).}
\label{hf2}
\end{center}
\end{figure}

The results shown in Fig. \ref{hf1} highlight the need to include
higher-order terms in the perturbative expansion of the
$\hat{Q}$-box.
In this connection, for the sake of completeness we compare the
results obtained in case (b) with those obtained without including
second- and third-order $(V-U)$-insertion diagrams in the
$\hat{Q}$-boxes, i.e. taking into account only the first-order one
(see Fig. \ref{Sbox1}).
In Fig. \ref{hf2}, the theoretical energies of the yrast states in
$^6$Li obtained with this procedure (case (c)) are reported and
compared with (b).

The results reported in Fig. \ref{hf2} show that without including
consistently order by order the $(V-U)$-insertion diagrams in the
$\hat{Q}$-box, a relevant dependence on $\hbar \omega$ may be
introduced.

Finally, it should be mentioned that in the works in
Refs. \cite{Ellis71,Kassis72}, the role of $(V-U)$-insertion diagrams
has been investigated.

\subsection{An application of the $\hat{Z}$-box graphical method:
  $(0+2)~\hbar \omega$ shell-model calculations}

Starting from a HO unperturbed hamiltonian, it is natural to extend
our calculations for the $p$-shell model space to a $(0+2)~\hbar
\omega$ one, which includes the $sd$-shell orbitals too.
In such a case, the unperturbed hamiltonian yields a non-degenerate
model space, the $p$ and $sd$ orbitals being separated by $\Delta
E=\hbar \omega$.

In Subsec. \ref{zbox} it has been shown that the $\hat{Z}$-box
graphical method provides a simple way to derive effective
hamiltonians for non-degenerate model spaces.
We have therefore found it interesting to employ it to obtain a
realistic effective shell-model hamiltonian for the $psd$ model space.

\begin{figure}[H]
\begin{center}
\includegraphics[scale=0.35,angle=0]{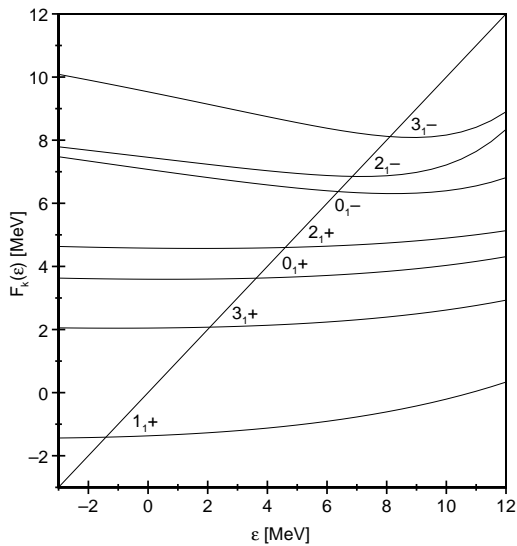}
\caption{$F_k(\epsilon)$ corresponding to $^{6}$Li yrast states, obtained with
  the N$^3$LO potential, as a function of the energy $\epsilon$ (see
  Eq. (\ref{eq-Z-eigenvalue})). The graph $y=\epsilon$ is also reported.}
\label{6LiZbox}
\end{center}
\end{figure}

To this end, we have calculated the Pad\`e approximant $[2|1]$ of
the $\hat{Q}$-box as a function of the energy $\epsilon$, starting
from the chiral N$^3$LO potential.
The $\hat{Q}(\epsilon)$ has been then employed to calculate the
vertex-function $\hat{Z}(\epsilon)$, as defined by
Eq. (\ref{eq:z-box}).
The $\hat{Z}(\epsilon)$ is the building block of the graphical method
we have described in Subsec. \ref{zbox} to derive $H^{\rm eff}$, whose
equations have been solved using the secant method algorithm (see
Fig. \ref{6LiZbox}).

We have to recall now that, when dealing with a $(0+2)~\hbar \omega$
model space, the results of the diagonalization of the shell-model
hamiltonian are affected by the spurious center-of-mass motion
\cite{Elliott55}.
The procedure we have followed to separate in energy the excitations
due to the internal degrees of freedom from those with spurious
center-of-mass components is the one suggested by Gloeckner and Lawson
in Ref. \cite{Gloeckner74}.

\begin{figure}[H]
\begin{center}
\includegraphics[scale=0.35,angle=0]{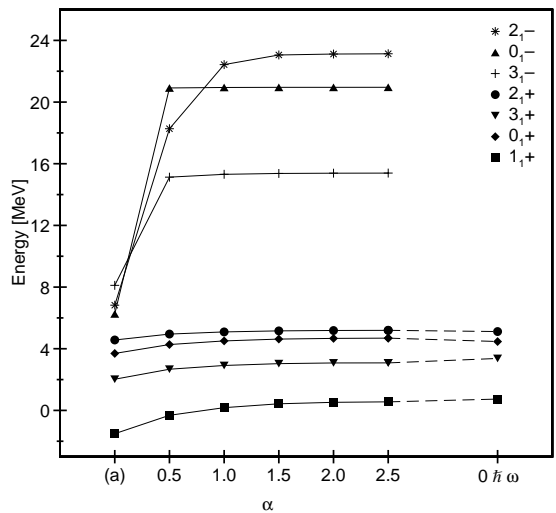}
\caption{Theoretical energies of $^{6}$Li yrast states, obtained with
  the N$^3$LO potential, relative to $^4$He as a function of parameter
  $\alpha=\log \beta$. (a) are results with $\beta=0$ (see text for
  details).}
\label{alpha}
\end{center}
\end{figure}

Thus we have diagonalized the modified shell-model hamiltonian $H'$

\begin{equation}
H' = H^{\rm eff} + H_{\beta} ~~,
\end{equation}

\noindent
where $H_{\beta}$ is $\beta$ times the center-of-mass excitation
energy of the $A$-nucleon system

\begin{equation} 
H_{\beta} = \beta \left\{ \frac{(\sum_{i=1}^A {\bf p_i})^2 }{2Am} +
\frac{m \omega^2}{2A} (\sum_{i=1}^A {\bf r_i})^2 -\frac{3}{2} \hbar
\omega \right\}~~.
\end{equation}

The spurious components are pushed up in energy by increasing the
parameter $\beta$, so that one can assume that the low-energy spectrum
is free from the above components.
Usually, the parameter $\beta$ is expressed as a power of ten
$\beta=10^{\alpha}$, and the hamiltonian $H'$ is diagonalized
increasing $\alpha$ until the low-energy eigenvalues are stable.

In Fig. \ref{alpha}, we report our results for the yrast states of
$^{6}$Li as a function of the $\alpha$ parameter, and compare the
positive parity spectrum with the one reported in Fig. \ref{N3LO3rd}
corresponding to the $0~\hbar \omega$ $p$-shell model space, using the
LS method where the Pad\`e approximant of the $\hat{Q}$-box has been
calculated.

From inspection of Fig. \ref{alpha}, it can be observed that the
results are quite stable for $\alpha \geq 1.5$.
Moreover, it should be noticed that the yrast $4^+$, $5^+$, $1^-$, and
$4^-$ states have not been reported since they turn out to be strongly
affected by center-of-mass spuriosity.

Finally, it is worth to note that yrast $0^+$, $1^+$, $2^+$, and
$3^+$ are in an excellent agreement with those calculated within the
$0~\hbar \omega$ model space.

\section{Comparison of realistic shell model with {\em ab initio}
  calculations}
\label{nocore}
In the previous subsections we have shown how the approximations
involved in the derivation of realistic shell-model effective
hamiltonians may be kept under control by way of some convergence
checks.

However, to study the accuracy of these approximations one should
compare the final results with those provided by an approach that
gives an ``exact'' solution of the Schr\"odinger equation (\ref{eq1}).

To this end, we will compare in this subsection our results obtained
for the $p$-shell nuclei starting from the N$^3$LO potential with
those provided from the {\it ab initio} no-core shell model (NCSM)
\cite{Navratil04,Navratil07a}.

All the results shown in the previous subsections have been obtained
starting from a $A$-body hamiltonian which is not translationally
invariant:

\begin{equation}
\label{ham}
H = \sum_{i=1}^A \frac{p_i^2}{2m} + \sum_{i < j =1}^A  V^{NN}_{ij}~~.
\end{equation}

In order to compare realistic shell model with NCSM we have to
employ a purely intrinsic Hamiltonian, so we have to remove the center
of mass kinetic energy from the hamiltonian of Eq. (\ref{ham}).

\begin{figure}[H]
\begin{center}
\includegraphics[scale=0.4,angle=0]{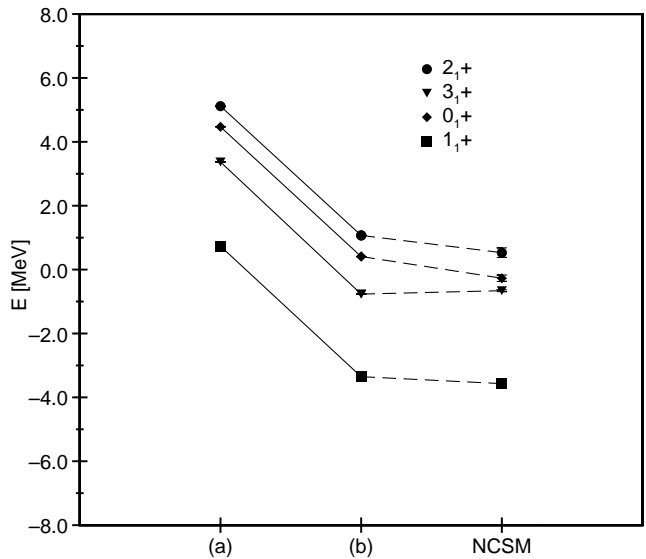}
\caption{Theoretical energies of $^{6}$Li yrast states relative to
  $^4$He, obtained with N$^3$LO potential. (a) shell-model calculation
  with an effective hamiltonian derived from Eq. (\ref{ham}). (b)
  shell-model calculation with an effective hamiltonian derived from
  Eq. (\ref{ham_int}). (c) NCSM calculation.}
\label{6Lincsm}
\end{center}
\end{figure}

We rewrite the intrinsic hamiltonian as follows:
\begin{eqnarray}
\label{ham_int}
H_{int} = &\left ( 1 - \frac{1}{A} \right ) \sum_{i=1}^A
\frac{p_i^2}{2m} + \sum_{i < j =1}^A \left ( V^{NN}_{ij} - \frac{{\bf
      p_i} \cdot {\bf p_j}}{mA} \right ) = & ~~ \\ \nonumber 
~ & \left[ \sum_{i=1}^A (\frac{p_i^2}{2m} + U_i) \right] +
\left[ \sum_{i < j =1}^A (V^{NN}_{ij} - U_i - \frac{p^2_i}{2mA} -
\frac{\bf p_i \cdot p_j}{mA}) \right] = & ~ \\ \nonumber 
~& H_0+H_1 ~~.~~~~~~~~~~~~~~~~~~~~~~~& ~~~
\end{eqnarray}
  
In Fig. \ref{6Lincsm} the calculated energies of the yrast states in
$^6$Li relative to $^4$He are reported. 
Results labelled with (a) refer to a shell-model calculation with an
effective hamiltonian derived from Eq. (\ref{ham}), while the spectrum
(b) corresponds to an effective hamiltonian derived from the
translationally invariant hamiltonian of Eq. (\ref{ham_int}).     
The NCSM spectrum is obtained using the value of the binding energy
as calculated in Ref. \cite{Navratil07a}, combined with the calculated
excitation energies reported in Ref. \cite{Navratil04}, and is
calculated with respect to the $^{4}$He ground state energy provided
by the N$^3$LO potential \cite{Navratil07b}. 

From inspection of Fig. \ref{6Lincsm}, it is evident the need
to employ a purely intrinsic hamiltonian when dealing with light
nuclei.
Moreover, it can be observed that the agreement between the
shell-model results and the NCSM ones is quite good.

\begin{figure}[H]
\begin{center}
\includegraphics[scale=0.5,angle=0]{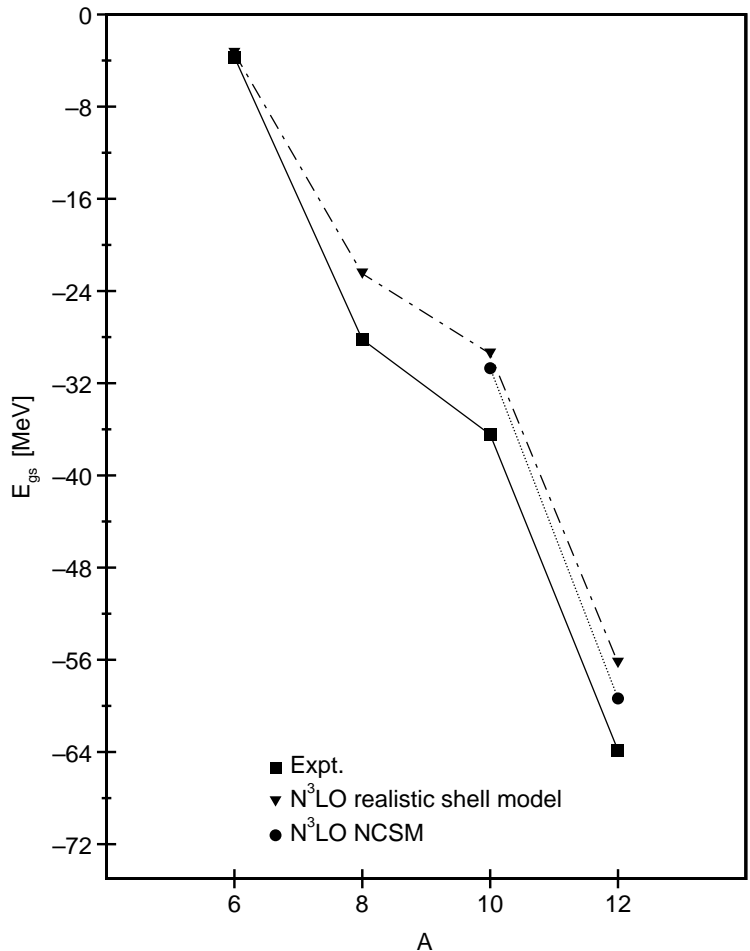}
\caption{Ground-state energies for $N=Z$ nuclei with mass $6 \leq A
  \leq 12 $.}
\label{gsencsm}
\end{center}
\end{figure}

At this point it is worth to remind that our shell-model effective
hamiltonian is derived for a system with only two valence-nucleons.
This implies that using this hamiltonian to study nuclei with $n>2$
valence nucleons, one introduces an approximation due to neglect the
3-, 4-, .. $n$-body components that arise in $H^{\rm eff}_1$, even if
the original hamiltonian contains only a two-body force.

With this remark in mind, we compare in Fig. \ref{gsencsm} the
ground-state energies, relative to $^4$He, for the $N=Z$ nuclei with
mass $6 \leq A \leq 12$ calculated within the framework of the
realistic shell model (dot-dashed line) and of the NCSM (dotted line),
with the experimental ones (continuous line) \cite{audi03}.
The shell-model ground state energy for $^6$Li is close to the one
derived with the NCSM as already shown in Fig. \ref{6Lincsm}, while
the discrepancy between the calculated values becomes more significant
going from $^{10}$B to $^{12}$C. 
This can be related to the many-body ($>2$) components of $H^{\rm
  eff}_1$, whose role grows with the number of valence nucleons.

\begin{figure}[H]
\begin{center}
\includegraphics[scale=0.38,angle=-90]{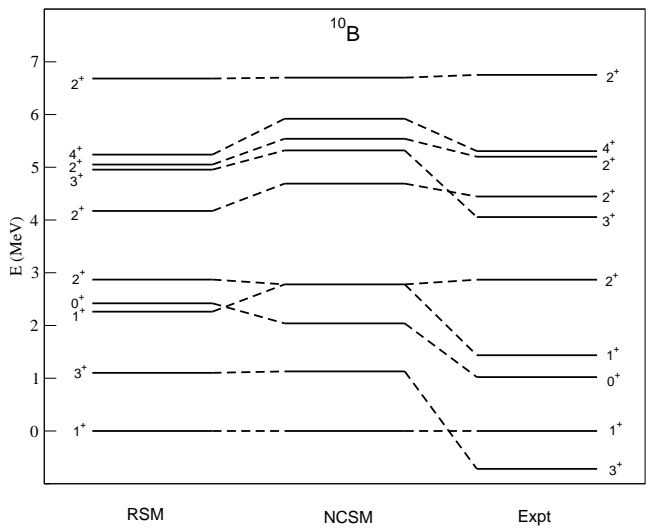}
\caption{Theoretical and experimental spectra for
  $^{10}$B. The theoretical energies have been obtained using the
  N$^3$LO potential within a realistic shell-model framework (RSM) and
  via a NCSM calculation (NCSM).}  
\label{B10ncsm}
\end{center}
\end{figure}

Fig. \ref{B10ncsm} shows the theoretical RSM and NCSM relative spectra
of $^{10}$B compared with the experimental one \cite{nndc}. 
From this figure it can be seen that the RSM excitation energies are
in good agreement with the NCSM ones, the largest discrepancy being
less than 1 MeV. 

It should be noted that in Ref. \cite{Navratil07a} it has been pointed
out how the contribution of a three-body force improves significantly
the reproduction by the theory of both ground state and excitation
energies.

\section{Summary and remarks}
This paper has been devoted to present the current status of the
derivation of a shell-model effective hamiltonian within the
perturbative approach, starting from a realistic $NN$ potential.

As we have already emphasized, the core of this approach is the
perturbative expansion of the well-known $\hat{Q}$-box vertex
function.
Therefore, we have focused our attention on the problems arising
from the calculation of the $\hat{Q}$-box perturbative series.
We have tackled different questions, all of them dealing with the
convergence properties (truncation of the intermediate-state space,
order-by-order convergence, dependence on the HO parameter).

At present, it is practically unfeasible to calculate contributions to
the perturbative expansion beyond the third order ones.
Thus we have presented results obtained by using the theory of the
Pad\`e approximants, starting from two different $NN$ chiral
potentials that are characterized by different offshell properties.
Comparing the results obtained with effective hamiltonians derived
calculating the $\hat{Q}$-box at third order in perturbation theory
and those obtained calculating the corresponding $[2|1]$ Pad\`e
approximant, we conclude that modern chiral $NN$ potentials provide
effective hamiltonians scarcely dependent on higher-order
contributions.

The foregoing conclusion is also borne out by the comparison of
shell-model results obtained with an effective hamiltonian derived
from the chiral N$^3$LO potential with those provided by {\em ab
  initio} NCSM calculations for some $p$-shell nuclei.

Finally, we have applied the recently introduced $\hat{Z}$-box
graphical method to derive a realistic effective hamiltonian for the
$(0+2)~\hbar \omega$ $psd$ model space.
The results show the versatility of this new method, which may be
easily applied to non-degenerate model spaces.

We hope that the discussions and results given in this paper may
provide useful guide to calculating shell-model realistic effective
hamiltonians.

\bibliographystyle{elsarticle-num}
\bibliography{biblio.bib}

%\newpage
\appendix
\section{Effective hamiltonians}
\label{effint}

\begin{table}[H]
\caption{Single-particle energies (in MeV) of different
  effective hamiltonians derived from the N$^3$LO potential and
  calculating the Pad\`e approximant $[2|1]$ of the
  $\hat{Q}$-box. Labels LS, $\hat{Z}$-box, and $A=6$ indicate SP
  energies obtained using the LS iterative technique, the
  $\hat{Z}$-box graphical method, and starting from a purely intrinsic
  hamiltonian for $A=6$ (see Sec. \ref{nocore}), respectively.}
\begin{center}
\begin{tabular}{crrrr} %{cccccc}
\hline
$nlj$ & $T_z$ & LS & $\hat{Z}$-box & $A=6$ \\
\hline
$0p_{3/2}$  &  1/2 & 4.593 & 4.543 & 2.747 \\
$0p_{1/2}$  &  1/2 & 7.449 & 7.512 & 5.886 \\
$0p_{3/2}$  & -1/2 & 3.915 & 3.904 & 2.052 \\
$0p_{1/2}$  & -1/2 & 7.077 & 7.163 & 5.720 \\
\hline
\end{tabular}
\end{center}
\label{spetabN3LO}
\end{table}

\begin{table}[H]
\caption{Same as in Table \ref{spetabN3LO} for the N$^3$LOW potential.}
\begin{center}
\begin{tabular}{crrrr} %{cccccc}
\hline
$nlj$ & $T_z$ & LS & $\hat{Z}$-box & $A=6$ \\
\hline
$0p_{3/2}$  &  1/2 & 4.941 & 4.925 & 3.083 \\
$0p_{1/2}$  &  1/2 & 7.647 & 7.755 & 5.655 \\
$0p_{3/2}$  & -1/2 & 4.001 & 3.987 & 2.126 \\
$0p_{1/2}$  & -1/2 & 6.954 & 6.892 & 5.186 \\
\hline
\end{tabular}
\end{center}
\label{spetabN3LOW}
\end{table}

\begin{table}[H]
\caption{Two-body matrix elements (in MeV) of different effective
  hamiltonians derived calculating the Pad\`e approximant $[2|1]$ of
  the $\hat{Q}$-box from the N$^3$LO potential. They are
  antisymmetrized, and normalized by a factor $1/ \sqrt{ (1 +
    \delta_{j_aj_b})(1 + \delta_{j_cj_d})}$. Labels are the same as in
  Table \ref{spetabN3LO}.}
\begin{center}
\begin{tabular}{ccrrrrr} %{ccccccc}
\hline
$n_a l_a j_a ~ n_b l_b j_b ~ n_c l_c j_c ~ n_d l_d j_d $ & $J$ & $T_z$
  &  LS &  $\hat{Z}$-box &  $A=6$ \\
\hline
 $ 0p_{ 3/2}~ 0p_{ 3/2}~ 0p_{ 3/2}~ 0p_{ 3/2}$ &  0 &  1 & -1.954 & -1.926 & -2.298 \\
 $ 0p_{ 3/2}~ 0p_{ 3/2}~ 0p_{ 1/2}~ 0p_{ 1/2}$ &  0 &  1 & -3.636 & -3.630 & -3.804 \\
 $ 0p_{ 1/2}~ 0p_{ 1/2}~ 0p_{ 1/2}~ 0p_{ 1/2}$ &  0 &  1 &  0.727 &  0.725 &  0.358 \\
 $ 0p_{ 3/2}~ 0p_{ 1/2}~ 0p_{ 3/2}~ 0p_{ 1/2}$ &  1 &  1 &  0.611 &  0.615 &  0.425 \\
 $ 0p_{ 3/2}~ 0p_{ 3/2}~ 0p_{ 3/2}~ 0p_{ 3/2}$ &  2 &  1 & -0.879 & -0.876 & -1.056 \\
 $ 0p_{ 3/2}~ 0p_{ 3/2}~ 0p_{ 3/2}~ 0p_{ 1/2}$ &  2 &  1 & -1.677 & -1.704 & -1.737 \\
 $ 0p_{ 3/2}~ 0p_{ 1/2}~ 0p_{ 3/2}~ 0p_{ 1/2}$ &  2 &  1 & -1.824 & -1.887 & -2.163 \\
 $ 0p_{ 3/2}~ 0p_{ 3/2}~ 0p_{ 3/2}~ 0p_{ 3/2}$ &  0 & -1 & -2.746 & -2.742 & -3.062 \\
 $ 0p_{ 3/2}~ 0p_{ 3/2}~ 0p_{ 1/2}~ 0p_{ 1/2}$ &  0 & -1 & -3.754 & -3.760 & -3.958 \\
 $ 0p_{ 1/2}~ 0p_{ 1/2}~ 0p_{ 1/2}~ 0p_{ 1/2}$ &  0 & -1 &  0.021 & -0.030 & -0.400 \\
 $ 0p_{ 3/2}~ 0p_{ 1/2}~ 0p_{ 3/2}~ 0p_{ 1/2}$ &  1 & -1 &  0.135 &  0.111 & -0.054 \\
 $ 0p_{ 3/2}~ 0p_{ 3/2}~ 0p_{ 3/2}~ 0p_{ 3/2}$ &  2 & -1 & -1.367 & -1.370 & -1.564 \\
 $ 0p_{ 3/2}~ 0p_{ 3/2}~ 0p_{ 3/2}~ 0p_{ 1/2}$ &  2 & -1 & -1.747 & -1.783 & -1.832 \\
 $ 0p_{ 3/2}~ 0p_{ 1/2}~ 0p_{ 3/2}~ 0p_{ 1/2}$ &  2 & -1 & -2.205 & -2.295 & -2.538 \\
 $ 0p_{ 3/2}~ 0p_{ 3/2}~ 0p_{ 3/2}~ 0p_{ 3/2}$ &  0 &  0 & -2.656 & -2.642 & -2.976 \\
 $ 0p_{ 3/2}~ 0p_{ 3/2}~ 0p_{ 1/2}~ 0p_{ 1/2}$ &  0 &  0 & -3.751 & -3.809 & -3.932 \\
 $ 0p_{ 1/2}~ 0p_{ 1/2}~ 0p_{ 1/2}~ 0p_{ 1/2}$ &  0 &  0 &  0.121 &  0.085 & -0.271 \\
 $ 0p_{ 3/2}~ 0p_{ 3/2}~ 0p_{ 3/2}~ 0p_{ 3/2}$ &  1 &  0 & -1.902 & -1.889 & -2.366 \\
 $ 0p_{ 3/2}~ 0p_{ 3/2}~ 0p_{ 3/2}~ 0p_{ 1/2}$ &  1 &  0 &  3.615 &  3.659 &  3.680 \\
 $ 0p_{ 3/2}~ 0p_{ 3/2}~ 0p_{ 1/2}~ 0p_{ 3/2}$ &  1 &  0 & -3.581 & -3.619 & -3.698 \\
 $ 0p_{ 3/2}~ 0p_{ 3/2}~ 0p_{ 1/2}~ 0p_{ 1/2}$ &  1 &  0 &  2.794 &  2.832 &  3.072 \\
 $ 0p_{ 3/2}~ 0p_{ 1/2}~ 0p_{ 3/2}~ 0p_{ 1/2}$ &  1 &  0 & -3.046 & -3.085 & -3.352 \\
 $ 0p_{ 3/2}~ 0p_{ 1/2}~ 0p_{ 1/2}~ 0p_{ 3/2}$ &  1 &  0 &  3.224 &  3.301 &  3.347 \\
 $ 0p_{ 3/2}~ 0p_{ 1/2}~ 0p_{ 1/2}~ 0p_{ 1/2}$ &  1 &  0 &  1.096 &  1.128 &  1.153 \\
 $ 0p_{ 1/2}~ 0p_{ 3/2}~ 0p_{ 1/2}~ 0p_{ 3/2}$ &  1 &  0 & -3.109 & -3.209 & -3.422 \\
 $ 0p_{ 1/2}~ 0p_{ 3/2}~ 0p_{ 1/2}~ 0p_{ 1/2}$ &  1 &  0 & -1.143 & -1.205 & -1.247 \\
 $ 0p_{ 1/2}~ 0p_{ 1/2}~ 0p_{ 1/2}~ 0p_{ 1/2}$ &  1 &  0 & -2.113 & -2.166 & -2.850 \\
 $ 0p_{ 3/2}~ 0p_{ 3/2}~ 0p_{ 3/2}~ 0p_{ 3/2}$ &  2 &  0 & -1.369 & -1.370 & -1.554 \\
 $ 0p_{ 3/2}~ 0p_{ 3/2}~ 0p_{ 3/2}~ 0p_{ 1/2}$ &  2 &  0 & -1.230 & -1.252 & -1.286 \\
 $ 0p_{ 3/2}~ 0p_{ 3/2}~ 0p_{ 1/2}~ 0p_{ 3/2}$ &  2 &  0 &  1.207 &  1.227 &  1.285 \\
 $ 0p_{ 3/2}~ 0p_{ 1/2}~ 0p_{ 3/2}~ 0p_{ 1/2}$ &  2 &  0 & -4.280 & -4.331 & -4.760 \\
 $ 0p_{ 3/2}~ 0p_{ 1/2}~ 0p_{ 1/2}~ 0p_{ 3/2}$ &  2 &  0 & -2.062 & -2.134 & -2.281 \\
 $ 0p_{ 1/2}~ 0p_{ 3/2}~ 0p_{ 1/2}~ 0p_{ 3/2}$ &  2 &  0 & -4.373 & -4.423 & -4.876 \\
 $ 0p_{ 3/2}~ 0p_{ 3/2}~ 0p_{ 3/2}~ 0p_{ 3/2}$ &  3 &  0 & -5.137 & -5.118 & -5.565 \\
\hline
\end{tabular}
\end{center}
\label{tableTBMEN3LO}
\end{table}

\begin{table}[H]
\caption{Same as in Table \ref{tableTBMEN3LO} for the
  N$^3$LOW potential.}
\begin{center}
\begin{tabular}{ccrrrrr} %{ccccccc}
\hline
$n_a l_a j_a ~ n_b l_b j_b ~ n_c l_c j_c ~ n_d l_d j_d $ & $J$ & $T_z$
  &  LS  &  $\hat{Z}$-box &  $A=6$ \\
\hline
 $ 0p_{ 3/2}~ 0p_{ 3/2}~ 0p_{ 3/2}~ 0p_{ 3/2}$ &  0 &  1 & -2.828 & -2.815 & -3.221 \\
 $ 0p_{ 3/2}~ 0p_{ 3/2}~ 0p_{ 1/2}~ 0p_{ 1/2}$ &  0 &  1 & -3.554 & -3.585 & -3.800 \\
 $ 0p_{ 1/2}~ 0p_{ 1/2}~ 0p_{ 1/2}~ 0p_{ 1/2}$ &  0 &  1 & -0.425 & -0.394 & -0.894 \\
 $ 0p_{ 3/2}~ 0p_{ 1/2}~ 0p_{ 3/2}~ 0p_{ 1/2}$ &  1 &  1 &  0.478 &  0.486 &  0.291 \\
 $ 0p_{ 3/2}~ 0p_{ 3/2}~ 0p_{ 3/2}~ 0p_{ 3/2}$ &  2 &  1 & -0.896 & -0.899 & -1.111 \\
 $ 0p_{ 3/2}~ 0p_{ 3/2}~ 0p_{ 3/2}~ 0p_{ 1/2}$ &  2 &  1 & -1.893 & -1.906 & -1.971 \\
 $ 0p_{ 3/2}~ 0p_{ 1/2}~ 0p_{ 3/2}~ 0p_{ 1/2}$ &  2 &  1 & -2.077 & -2.109 & -2.447 \\
 $ 0p_{ 3/2}~ 0p_{ 3/2}~ 0p_{ 3/2}~ 0p_{ 3/2}$ &  0 & -1 & -3.573 & -3.554 & -3.892 \\
 $ 0p_{ 3/2}~ 0p_{ 3/2}~ 0p_{ 1/2}~ 0p_{ 1/2}$ &  0 & -1 & -3.779 & -3.788 & -3.995 \\
 $ 0p_{ 1/2}~ 0p_{ 1/2}~ 0p_{ 1/2}~ 0p_{ 1/2}$ &  0 & -1 & -1.033 & -1.014 & -1.477 \\
 $ 0p_{ 3/2}~ 0p_{ 1/2}~ 0p_{ 3/2}~ 0p_{ 1/2}$ &  1 & -1 &  0.042 &  0.042 & -0.161 \\
 $ 0p_{ 3/2}~ 0p_{ 3/2}~ 0p_{ 3/2}~ 0p_{ 3/2}$ &  2 & -1 & -1.399 & -1.400 & -1.620 \\
 $ 0p_{ 3/2}~ 0p_{ 3/2}~ 0p_{ 3/2}~ 0p_{ 1/2}$ &  2 & -1 & -1.985 & -1.987 & -2.064 \\
 $ 0p_{ 3/2}~ 0p_{ 1/2}~ 0p_{ 3/2}~ 0p_{ 1/2}$ &  2 & -1 & -2.478 & -2.498 & -2.791 \\
 $ 0p_{ 3/2}~ 0p_{ 3/2}~ 0p_{ 3/2}~ 0p_{ 3/2}$ &  0 &  0 & -3.523 & -3.508 & -3.862 \\
 $ 0p_{ 3/2}~ 0p_{ 3/2}~ 0p_{ 1/2}~ 0p_{ 1/2}$ &  0 &  0 & -3.743 & -3.759 & -3.957 \\
 $ 0p_{ 1/2}~ 0p_{ 1/2}~ 0p_{ 1/2}~ 0p_{ 1/2}$ &  0 &  0 & -0.987 & -0.966 & -1.436 \\
 $ 0p_{ 3/2}~ 0p_{ 3/2}~ 0p_{ 3/2}~ 0p_{ 3/2}$ &  1 &  0 & -2.283 & -2.254 & -2.676 \\
 $ 0p_{ 3/2}~ 0p_{ 3/2}~ 0p_{ 3/2}~ 0p_{ 1/2}$ &  1 &  0 &  3.631 &  3.626 &  3.682 \\
 $ 0p_{ 3/2}~ 0p_{ 3/2}~ 0p_{ 1/2}~ 0p_{ 3/2}$ &  1 &  0 & -3.542 & -3.536 & -3.661 \\
 $ 0p_{ 3/2}~ 0p_{ 3/2}~ 0p_{ 1/2}~ 0p_{ 1/2}$ &  1 &  0 &  2.737 &  2.759 &  3.130 \\
 $ 0p_{ 3/2}~ 0p_{ 1/2}~ 0p_{ 3/2}~ 0p_{ 1/2}$ &  1 &  0 & -3.462 & -3.439 & -3.709 \\
 $ 0p_{ 3/2}~ 0p_{ 1/2}~ 0p_{ 1/2}~ 0p_{ 3/2}$ &  1 &  0 &  3.486 &  3.464 &  3.520 \\
 $ 0p_{ 3/2}~ 0p_{ 1/2}~ 0p_{ 1/2}~ 0p_{ 1/2}$ &  1 &  0 &  1.183 &  1.197 &  1.098 \\
 $ 0p_{ 1/2}~ 0p_{ 3/2}~ 0p_{ 1/2}~ 0p_{ 3/2}$ &  1 &  0 & -3.444 & -3.423 & -3.656 \\
 $ 0p_{ 1/2}~ 0p_{ 3/2}~ 0p_{ 1/2}~ 0p_{ 1/2}$ &  1 &  0 & -1.273 & -1.288 & -1.254 \\
 $ 0p_{ 1/2}~ 0p_{ 1/2}~ 0p_{ 1/2}~ 0p_{ 1/2}$ &  1 &  0 & -3.277 & -3.264 & -3.918 \\
 $ 0p_{ 3/2}~ 0p_{ 3/2}~ 0p_{ 3/2}~ 0p_{ 3/2}$ &  2 &  0 & -1.396 & -1.400 & -1.616 \\
 $ 0p_{ 3/2}~ 0p_{ 3/2}~ 0p_{ 3/2}~ 0p_{ 1/2}$ &  2 &  0 & -1.393 & -1.396 & -1.453 \\
 $ 0p_{ 3/2}~ 0p_{ 3/2}~ 0p_{ 1/2}~ 0p_{ 3/2}$ &  2 &  0 &  1.372 &  1.375 &  1.461 \\
 $ 0p_{ 3/2}~ 0p_{ 1/2}~ 0p_{ 3/2}~ 0p_{ 1/2}$ &  2 &  0 & -4.678 & -4.693 & -5.098 \\
 $ 0p_{ 3/2}~ 0p_{ 1/2}~ 0p_{ 1/2}~ 0p_{ 3/2}$ &  2 &  0 & -2.055 & -2.041 & -2.285 \\
 $ 0p_{ 1/2}~ 0p_{ 3/2}~ 0p_{ 1/2}~ 0p_{ 3/2}$ &  2 &  0 & -4.675 & -4.690 & -5.090 \\
 $ 0p_{ 3/2}~ 0p_{ 3/2}~ 0p_{ 3/2}~ 0p_{ 3/2}$ &  3 &  0 & -5.569 & -5.568 & -6.014 \\
\hline
\end{tabular}
\end{center}
\label{tableTBMEN3LOW}
\end{table}

\section{Diagrammatics}
\label{graphs}

\begin{figure}[H]
\begin{center}
\includegraphics[scale=0.5,angle=0]{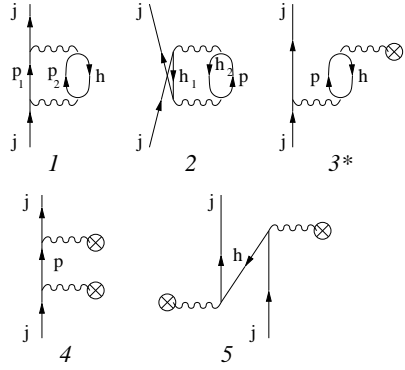}
\caption{One-body second-order diagrams. The asterisk indicates
  non-symmetric diagrams, which occur always in pairs giving equal
  contributions. For the sake of simplicity we report only one of
  them.}
\label{Sbox2}
\end{center}
\end{figure}

\begin{figure}[H]
\begin{center}
\includegraphics[scale=0.85,angle=0]{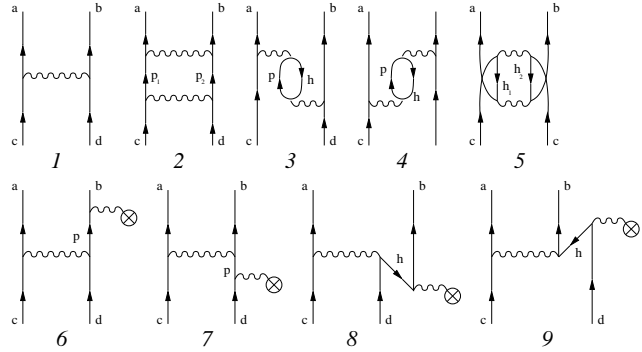}
\caption{Two-body diagrams up to second order in perturbation
  theory. For the sake of simplicity, for each topology we report only
  one of the diagrams which correspond to the exchange of the external
  pairs of lines.}
\label{Qbox2}
\end{center}
\end{figure}

\begin{figure}[H]
\begin{center}
\includegraphics[scale=0.75,angle=0]{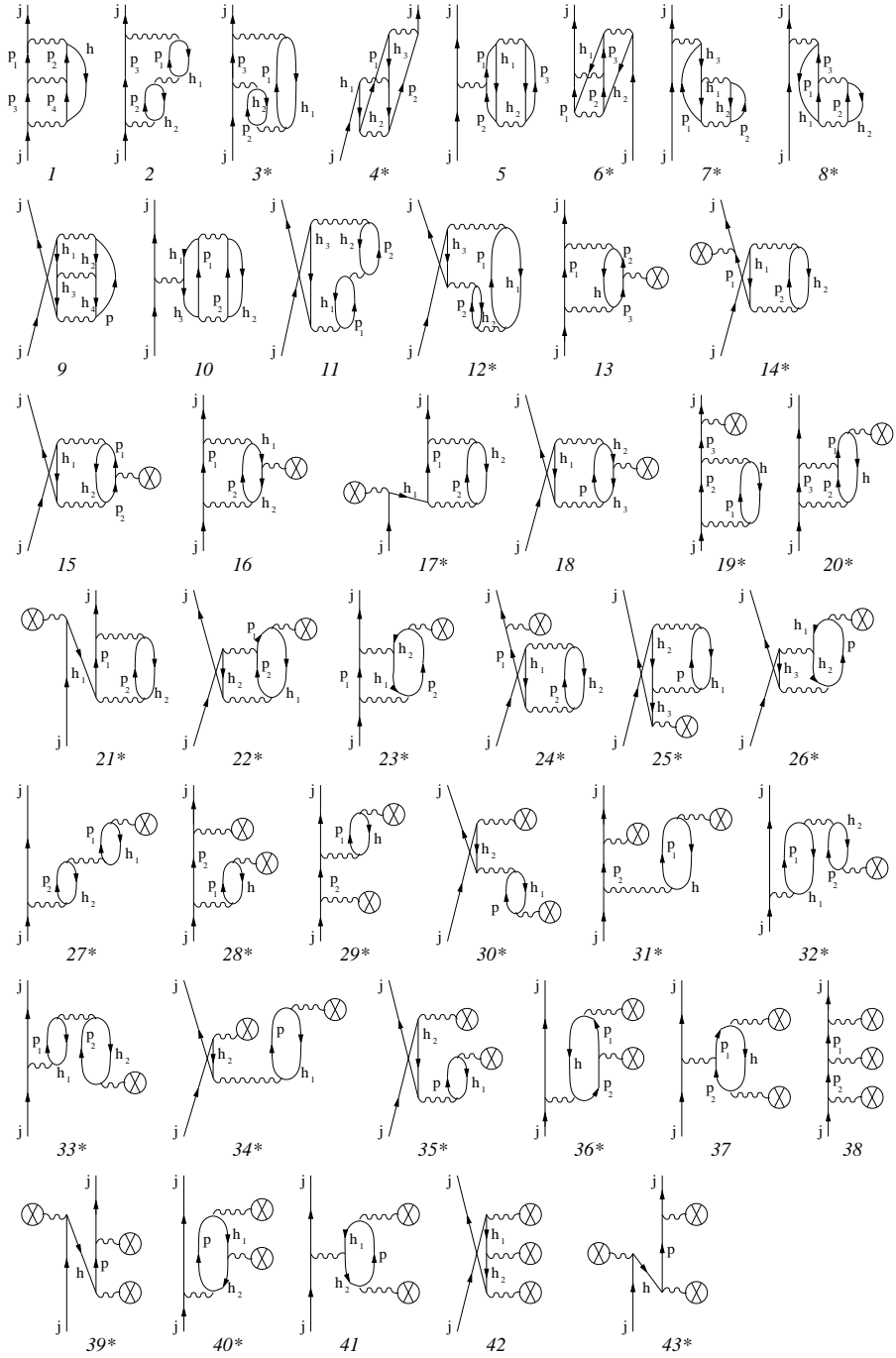}
\caption{Same as Fig. \ref{Sbox2}, but for third-order diagrams.}
\label{Sbox3}
\end{center}
\end{figure}

\begin{figure}[H]
\begin{center}
\includegraphics[scale=0.7,angle=0]{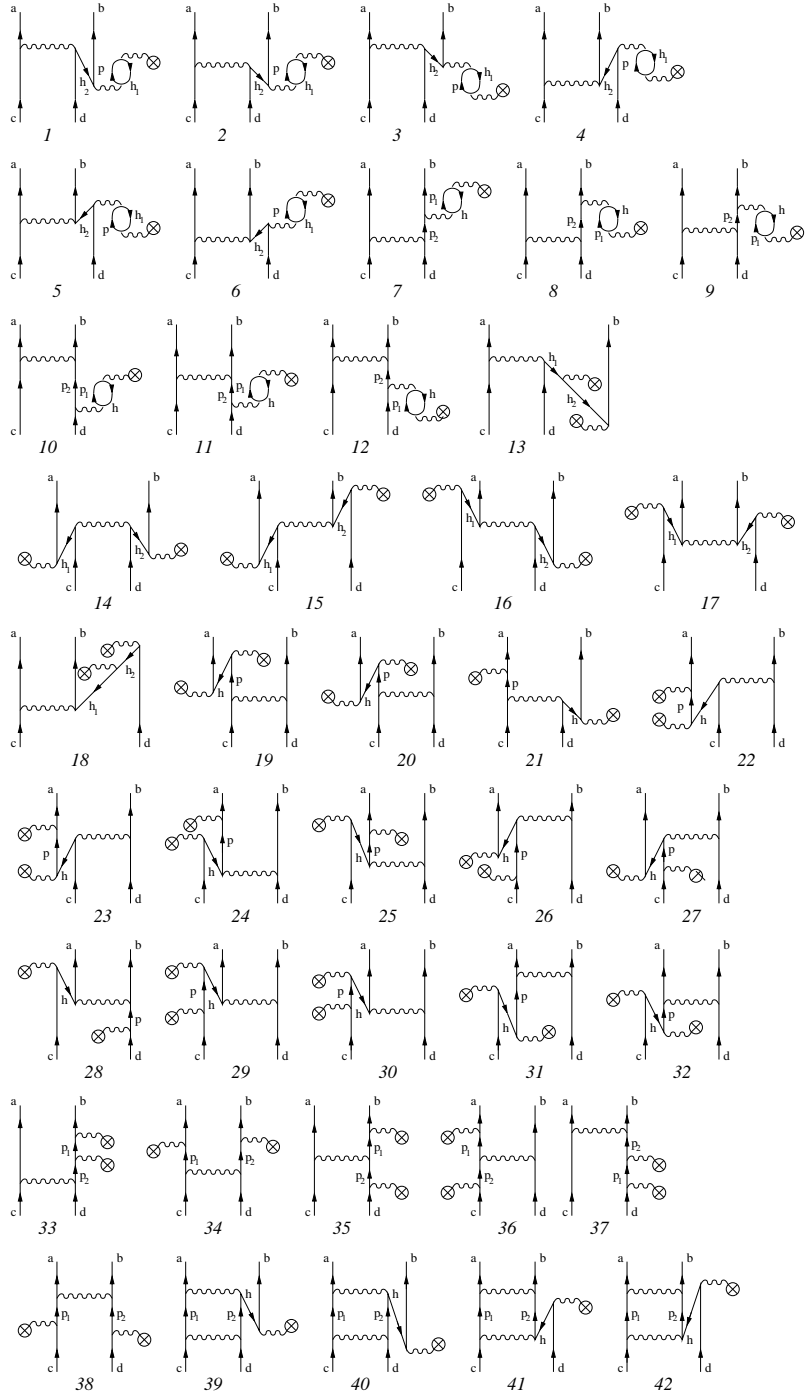}
\caption{Same as Fig. \ref{Qbox2}, but for third-order diagrams.}
\label{Qbox3a}
\end{center}
\end{figure}

\begin{figure}[H]
\begin{center}
\includegraphics[scale=0.7,angle=0]{Q-box_3_b.epsi}
\caption{Same as Fig. \ref{Qbox3a}.}
\label{Qbox3b}
\end{center}
\end{figure}

\begin{figure}[H]
\begin{center}
\includegraphics[scale=0.75,angle=0]{Q-box_3_c.epsi}
\caption{Same as Fig. \ref{Qbox3a}.}
\label{Qbox3c}
\end{center}
\end{figure}

\newpage

%% The Appendices part is started with the command \appendix;
%% appendix sections are then done as normal sections
%% \appendix

%% \section{}
%% \label{}

%% References
%%
%% Following citation commands can be used in the body text:
%% Usage of \cite is as follows:
%%   \cite{key}         ==>>  [#]
%%   \cite[chap. 2]{key} ==>> [#, chap. 2]
%%

%% References with bibTeX database:

%% Authors are advised to submit their bibtex database files. They are
%% requested to list a bibtex style file in the manuscript if they do
%% not want to use elsarticle-num.bst.

%% References without bibTeX database:

% \begin{thebibliography}{00}

%% \bibitem must have the following form:
%%   \bibitem{key}...
%%

% \bibitem{}

% \end{thebibliography}

\end{document}